\documentclass[aps,amssymb,amsmath,pra,superscriptaddress]{revtex4}
\usepackage{graphicx}
\usepackage{epsfig}
\usepackage{dcolumn}
\usepackage{bm}
\usepackage{bbm}
\usepackage[up]{subfigure}
\newtheorem{thm}{Theorem}

\newcommand{\be}{\begin{equation}}
\newcommand{\ee}{\end{equation}}
\newcommand{\bc}{\begin{center}}
\newcommand{\ec}{\end{center}}
\newcommand{\bea}{\begin{eqnarray}}
\newcommand{\eea}{\end{eqnarray}}

\begin{document}
\title{Symmetries and noise in quantum walk}
\author{C. M. \surname{Chandrashekar}}
\email{cmadaiah@iqc.ca}
\affiliation{Institute for Quantum Computing, University of Waterloo, Canada}
\author{R. \surname{Srikanth}}
\email{srik@rri.res.in}
\affiliation{Poornaprajna Institute of Scientific Research,
Devanahalli, Bangalore 562 110, India}
\affiliation{Raman Research Institute, Sadashiva Nagar, Bangalore, India}
\author{Subhashish Banerjee}
\email{subhashishb@rri.res.in}
\affiliation{Raman Research Institute, Sadashiva Nagar, Bangalore, India}

\begin{abstract}
We study  some discrete symmetries  of unbiased (Hadamard)  and biased
quantum walk on a line, which  are shown to hold even when the quantum
walker  is  subjected  to  environmental effects.   The  noise  models
considered in order  to account for these effects  are the phase flip,
bit flip  and generalized  amplitude damping channels.   The numerical
solutions  are  obtained  by  evolving  the density  matrix,  but  the
persistence of the symmetries in the presence of noise is proved using
the  quantum  trajectories  approach.  We also  briefly  extend  these
studies  to quantum  walk on  a  cycle.  These  investigations can  be
relevant  to the  implementation  of quantum  walks  in various  known
physical systems.   We discuss the  implementation in the case  of NMR
quantum information processor and ultra cold atoms.
\end{abstract}


\maketitle
\preprint{Version}

\section{Introduction}
\label{sec:intro}

Random   walk,   which  has   found   applications   in  many   fields
\cite{barber-ninham,   chandra},  is   an  important   constituent  of
information theory. It  has played a very prominent  role in classical
computation.  Markov chain simulation, which has emerged as a powerful
algorithmic  tool  \cite{mark},  as   well  as  many  other  classical
algorithms,  are   based  on  random  walks.    Quantum  random  walks
\cite{aharonov},  the  generalization  of  classical random  walks  to
situations where  the quantum  uncertainties play a  predominant role,
are  of both  mathematical  and experimental  interest  and have  been
investigated by  a number  of groups.  It  is believed  that exploring
quantum random walks  \cite{kempe} allows, in a similar  way, a search
for new quantum algorithms, a  few of which have already been proposed
\cite{childs, shenvi, childs1, ambainis}.  Experimental implementation
of the quantum  walk has also been reported  \cite{ryan, du}.  Various
other schemes have  been proposed for the physical  realization of the
quantum walks \cite{travaglione, rauss, eckert, chandra06, ma}.

The evolution of a discrete  classical random walk, involving steps of
a given length, is described  in terms of probabilities.  On the other
hand, the evolution of a  discrete quantum random walk is described in
terms  of   probability  amplitudes.   An   unbiased  one  dimensional
classical random  walk with the particle initially  at $x_{0}$ evolves
in such a  way that at each step, the  particle moves with probability
1/2 one step to the left  or right. In a quantum mechanical analog the
state  of  the   particle  evolves  at  each  step   into  a  coherent
superposition of  moving one step to  the right {\em and}  one step to
the left.

In the one dimensional  quantum (Hadamard) walk the particle initially
prepared  in a  product  state  of the  coin  (internal) and  position
(external) degree of  freedom, is subjected to a  rotation in the coin
Hilbert space, followed by a conditional shift operation, to evolve it
into a  superposition in the  position space. The process  is iterated
without  resorting to  intermediate  measurements to  realize a  large
number of steps of quantum walk, before a final measurement.

\begin{figure}
\begin{center}
\epsfig{figure=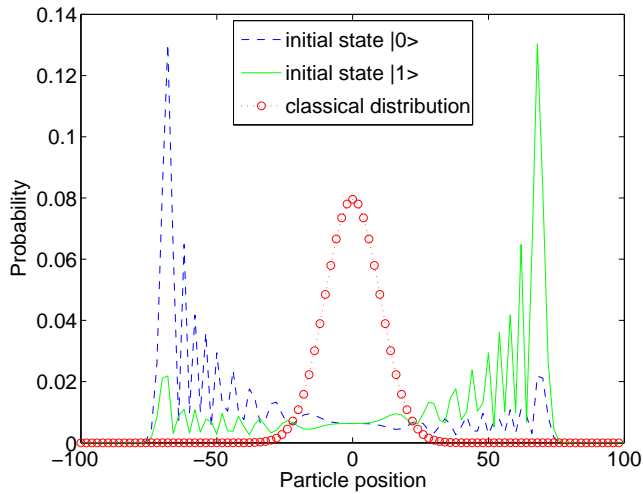, height= 7.00cm}
\caption{\label{u} (color online) Probability  distribution of the  quantum walk with
the  initial   state  $|0\rangle$  (solid line)   and  initial  state
$|1\rangle$  (dashed line) on  the position.  
The distribution is for 100 steps.}
\end{center}
\end{figure}

\begin{figure}
\begin{center}
\epsfig{figure=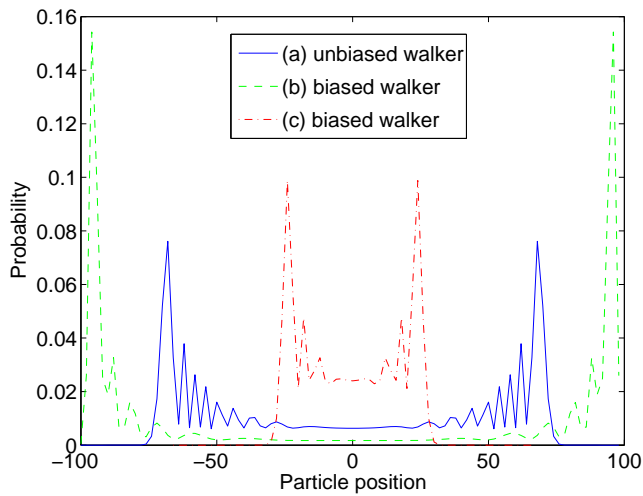, height= 7.00cm}
\caption{\label{qw}  (color online) The  probability   distribution  of  an  unbiased
walker, and  a biased  quantum walker using  an SU(2) operator  of the
form (\ref{eq:bias1}) as the  quantum coin toss.  (a) Unbiased quantum
walker ($\theta = 45^{\circ}$); (b) Biased quantum walker with $\theta
=   15^{\circ}$;   (c)   Biased   quantum  walker   with   $\theta   =
75^{\circ}$. The distribution is for 100 steps.}
\end{center}
\end{figure}

A one dimensional quantum  (Hadamard) walk starting from initial state
$|0\rangle \otimes |x_{0}\rangle$ or $|1\rangle \otimes |x_{0}\rangle$
(where the  first register refers to  the coin degree  of freedom, and
the  second  register to  the  external,  spatial  degree of  freedom)
induces an asymmetric probability distribution of finding the particle
after $N$  number of  steps of quantum  walk. A particle  with initial
coin  state $|0\rangle$  drifts to  the  right (solid  line in  Figure
\ref{u}) and particle with initial state $|1\rangle$ drifts to the left
(dashed line in  Figure \ref{u}). This asymmetry arises  from the fact
that  the Hadamard  coin  treats the  two  directions $|0\rangle$  and
$|1\rangle$ differently; it multiplies the phase by -1 only in case of
$|0\rangle$.  It follows that to obtain left-right symmetry, one needs
to  start the  coin  in  the state  $|0\rangle  + i|1\rangle$  (Figure
\ref{qw}).

In  this paper we  report that  the quantum  walk-- both  the unbiased
(Hadamard)  as well as  the biased--  remains invariant  under certain
operations,  i.e.,  the   probability  distribution  of  the  walker's
position remains unchanged under  the inclusion of these operations at
each  step of  the  walk. We  refer  to these  discrete operations  as
symmetries of the  quantum walk. We further study  these symmetries in
the  presence  of  environmental  effects, modeled  by  various  noise
channels.  These results were obtained by numerical integration rather
than simulation.  We extend these  studies to quantum walk on a cycle,
which can  be conveniently generalized  to more general graphs.  It is
shown that  the above symmetries do not hold, in general, for a quantum  
walk on a
cycle and hence  for the closed graph, but  leads to other interesting
behavior.  These  observations can  have important implications  for a
better insight  into, and for simplifying  certain implementations of,
quantum walks.

This  paper is  organized as  follows.  Section  \ref{sec:qrw} briefly
recapitulates   the   theory   of   quantum  random   walk.    Section
\ref{sec:symm}  discusses quantum walk  augmented by  various symmetry
operations, both  in the case of  a biased and  unbiased quantum coin.
The above observations are generalized  to the case of a noisy quantum
walk  in Section \ref{sec:env},  with Section  \ref{sec:enva} treating
noise as phase-flip and  bit-flip channels, and Section \ref{sec:envb}
treating noise  as a  generalized amplitude damping  channel.  Whereas
the  numerical results  presented here,  involve evolving  the density
operator of  the system,  we have found  it convenient to  explain the
symmetries using quantum  trajectories.  In Section \ref{sec:cycle} we
extend these  studies to the  quantum walk on  the cycle which  can be
generalzed to all closed graphs in general. In Section \ref{sec:qwbec}
we show that the application  of these ideas can help simplify certain
experimental  implementations of  quantum  walk. We  also discuss  the
physical  systems,  NMR  quantum-information processor  and  ultracold
atoms, where the results presented  in this article can be applied. In
Section \ref{conclusion}, we make our conclusions.

\section{Quantum random walk}\label{sec:qrw}

Unlike classical random walk, two  degrees of freedom are required for
quantum  random walks,  the internal,  {\em coin}  degree  of freedom,
represented  by the  two-dimensional Hilbert  space  $\mathcal H_{c}$,
spanned  by the  basis  states $|0\rangle$  and  $|1\rangle$, and  the
particle degree of freedom,  represented by the {\it position} Hilbert
space $\mathcal  H_{p}$, spanned by  basis states $|x\rangle$,  $x \in
\mathbb{Z}$.  The state of the  total system is in the space $\mathcal
H= \mathcal H_{c}  \otimes \mathcal H_{p}$. The internal  state of the
particle determines the direction  of the particle's movement when the
conditional  unitary shift operator  $U$ is  applied on  the particle,
whose  initial  state  is  given  by a  product  state,  for  example,
$|\Psi_{in}\rangle=\frac{1}{\sqrt
2}[|0\rangle+|1\rangle]\otimes|\Psi_{x_{0}}\rangle$, where
\begin{eqnarray}
\label{eq:condshift}
U &=&|0\rangle \langle  0|\otimes \sum_{x \in \mathbb{Z}}
|x-1\rangle \langle x |+|1\rangle  \langle 1 |\otimes \sum_{x \in \mathbb{Z}} |x+1\rangle
\langle x| \nonumber \\
&\equiv & |0\rangle\langle0|\otimes \hat{A} + |1\rangle\langle1|
\otimes \hat{A}^{\dag}.
\end{eqnarray}
Here $\hat{A}$ and $\hat{A}^{\dag}$ are unitary operators that are
notationally reminiscent of annihilation and creation operations,
respectively. The conditional shift can also be written as 
\be
\label{eq:alter}
U = \exp(-2iS_{z}\otimes Pl),
\ee
\noindent
$P$, being the momentum operator and $S_{z}$, the operator corresponding
to the step of length $l$.

Conditioned on the internal state being $|0\rangle$ ($|1\rangle$), the
particle moves to the left (right), i.e.,
$U|0\rangle\otimes|x\rangle=|0\rangle\otimes|x-1\rangle$            and
$U|1\rangle\otimes|x\rangle=|1\rangle\otimes|x+1\rangle$.  Application
of $U$ on $|\Psi_{in}\rangle$ spatially entangles the $\mathcal H_{c}$
and  $\mathcal  H_{p}$ and  implements  quantum  (Hadamard) walk,  
\begin{equation}
\label{eq:uprim}
U|\Psi_{in}\rangle=\frac{1}{\sqrt     2}[|0\rangle\otimes     e^{-iPl}
+|1\rangle\otimes  e^{iPl}]|\Psi_{x_{0}}\rangle. 
\end{equation}
Each step of the quantum (Hadamard) walk is composed of a
Hadamard operation (rotation) $H$,
\begin{equation}
H =\frac{1}{\sqrt 2} \left( \begin{array}{clcr}
 1  & &   1   \\
 1  & &  -1
 \end{array} \right),
\end{equation}
\noindent
on  the particle, bringing  them to a superposition  state with  equal
probability,       such      that,       
\begin{eqnarray} 
\label{hadamard} 
(H\otimes\mathbbm{1})|0,x\rangle &=& 
\frac{1}{\sqrt 2}[|0,x\rangle+|1, x\rangle], \nonumber \\
(H\otimes \mathbbm{1})|1,  x\rangle &=& 
\frac{1}{\sqrt 2}[|0, x\rangle-|1,
x\rangle],
\end{eqnarray}
and, a subsequent unitary conditional
shift operation, $U$, which  moves the particle
into  an entangled state in  the  position space.   

The  probability  amplitude  distribution  arising from  the  iterated
application of $W=U(H\otimes  \mathbbm{1})$ is significantly different
from  the distribution  of  the  classical walk  after  the first  two
steps~\cite{kempe}.   If  the   coin  initially   is  in   a  suitable
superposition  of  $|0\rangle$ and  $|1\rangle$  then the  probability
amplitude distribution after  $n$ steps of quantum walk  will have two
maxima symmetrically displaced from  the starting point.  The variance
of quantum version is known to grow quadratically with number of steps
$n$,  $\sigma^{2}\propto   n^{2}$  compared  to   the  linear  growth,
$\sigma^{2}\propto n$ for the classical walk.

Figure  \ref{qw} shows  the  probability distribution  of finding  the
position of  the quantum walker using  a biased walker  with $\theta =
15^{\circ}$ and  $\theta = 75^{\circ}$, respectively,  using a general
$SU(2)$ operator of the form,
\begin{equation}
B(\xi,\theta,\zeta) =\left( \begin{array}{clcr}
 e^{i\xi}\cos(\theta)  & &   e^{i\zeta}\sin(\theta)   \\
e^{-i\zeta} \sin(\theta)  & &  -e^{-i\xi}\cos(\theta)
 \end{array} \right),
\label{eq:bias1}
\end{equation}
where we  have set $\xi=\zeta=0$.  Note that  $H = B(0,45^{\circ},0)$.
Replacing the Hadamard coin  with an arbitrary $SU(2)$ operator yields
a  biased  coin  toss,  with   $\theta  <  45^{\circ}$  or  $\theta  >
45^{\circ}$.  The  value  of   $\sigma$  can  be  increased  or  decreased,
respectively.

\section{Bit flip and phase flip symmetries in quantum walk}
\label{sec:symm}

Consider the application of the modified conditional shift operator
$U^{\prime}=|1\rangle  \langle  0|\otimes  \hat{A}
+|0\rangle  \langle  1 |\otimes \hat{A}^{\dag}$
instead of Eq. (\ref{eq:condshift}). In place of Eq. (\ref{eq:alter}),
one      has:     
\begin{equation}      U^{\prime}      =     (X      \otimes
\mathbb{I})\exp(-2iS_{z}\otimes Pl), 
\end{equation}  
where $X$ is  the Pauli $X$ operator.  Since  $U^{\prime} = XU$, i.e.,
it  is  equivalent  to  an  application of  bit  flip  following  $U$,
conditioned on the internal state being $|0\rangle$ ($|1\rangle$), the
particle will move to the  left (right) and changes its internal state
to       $|1\rangle$      ($|0\rangle$).        Thus,      $U^{\prime}
|0\rangle\otimes|x\rangle=|1\rangle\otimes|x-1\rangle$              and
$U^{\prime}|1\rangle\otimes|x\rangle=|0\rangle\otimes|x+1\rangle$.

A  relevant observation  in this  context is  that there  are physical
systems where  the implementation of $U^{\prime}$ is  easier than that
of  $U$ \cite{chandra06}. In  that case,  applying a  compensatory bit
flip on  the internal state,  after each application  of $U^{\prime}$,
reduces the  modified quantum  walk to the  usual scheme. In  all this
would  require  $(n-1)$ compensatory  bit  flips,  which  adds to  the
complexity of the  experimental realization.  However, this additional
complexity can be eliminated. For an unbiased quantum walk, applying a
bit  flip in each  step can  be shown  to be  equivalent to  a spatial
inversion of  the position probability  distribution.  A quick  way to
see why  bit flips (for an  unbiased quantum walk) are  harmless is to
note  that they are  also equivalent  to relabeling  the edges  of the
graph on which the quantum walk  takes place, so that each end of each
edge has  the same  label \cite{ken05}.  (It  is worth noting  that in
Ref.  \cite{rauss}, bit flips  are employed  to improve  the practical
implementation of a quantum walk  of atoms in an optical lattice.)  We
may in  this sense call a  bit flip together with  spatial inversion a
{\em  symmetry}  of  the unbiased  quantum  walk  on  a line.   To  be
specific, a quantum  walk symmetry is any unitary  operation which may
uniformly augment  each step of  a quantum walk without  affecting the
position probability distribution.  Experimentally, the symmetries are
useful in identifying  variants of a given quantum  walk protocol that
are equivalent to it.  This  motivates us to look for other (discrete)
symmetries of the  quantum walk, which we study  below.  We begin with
Theorem  \ref{thm:bias},  where  we  note  four  discrete  symmetries,
associated     with    the     matrices     $B^{(j)}$    $(j=1,2,3,4)$
(cf. Eq.  (\ref{eq:bias20})), of the  quantum walk. Thereafter  two of
these  symmetries,  $B^{(1)}$   and  $B^{(2)}$,  are  identified  with
operations  that  are  relevant   from  the  perspective  of  physical
implementation. It is an interesting open question of relevance to practical
implementation of quantum walks, whether other such symmetries of the
quantum walk exist.

\begin{thm} If $B$ in Eq. (\ref{eq:bias1}) is replaced
by any of $B^{(1)}$, $B^{(2)}$, 
$B^{(3)}$, or $B^{(4)}$, given by,
\begin{eqnarray}
B^{(1)} \equiv \left( \begin{array}{clcr}
 e^{i\xi}\cos(\theta)  & &   e^{i\zeta}\sin(\theta)   \\
 e^{i(\phi-\zeta)}\sin(\theta)  & & -e^{i(\phi-\xi)}\cos(\theta)
 \end{array} \right),\hspace{0.4cm} &&
B^{(2)} \equiv \left( \begin{array}{clcr}
 e^{i\xi}\cos(\theta)  & & e^{i(\phi+\zeta)}\sin(\theta)   \\
e^{-i\zeta}\sin(\theta)  & & -e^{i(\phi-\xi)}\cos(\theta)
 \end{array} \right),\nonumber \\
B^{(3)} \equiv \left( \begin{array}{clcr}
e^{i(\phi+\xi)}\cos(\theta)  & & e^{i(\phi+\zeta)}\sin(\theta)   \\
e^{-i\zeta}\sin(\theta)  & & -e^{-i\xi}\cos(\theta)
 \end{array} \right),\hspace{0.4cm} &&
B^{(4)} \equiv \left( \begin{array}{clcr}
e^{i(\phi+\xi)}\cos(\theta)  & & e^{i\zeta}\sin(\theta)   \\
e^{i(\phi-\zeta)}\sin(\theta)  & & -e^{-i\xi}\cos(\theta)
 \end{array} \right),
\label{eq:bias20}
\end{eqnarray}
the resulting position probability
distribution of the quantum walk remains unchanged.
\label{thm:bias}
\end{thm}

\noindent {\bf  Proof.}  With the notation $B  \equiv \{b_{j,k}\}$ and
$B^{(j)}   \equiv   \{b^{(j)}_{j,k}\}$,   we  find   $b^{(1)}_{jk}   =
b_{jk}e^{ij\phi}$, $b^{(2)}_{jk}  = b_{jk}e^{ik\phi}$, $b^{(3)}_{jk} =
b_{jk}e^{i\bar{j}\phi}$,   $b^{(4)}_{jk}   =  b_{jk}e^{i\bar{k}\phi}$,
where  the  matrix  indices $j,k$  take  values  0  and 1,  $i  \equiv
+\sqrt{-1}$,   and   the  overbar   denotes   a   NOT  operation   ($0
\leftrightarrow  1$).  The  state  vector obtained,  after $n$  steps,
using  $B$  and  $B^{(1)}$  as  the  coin  rotation  operations,  are,
respectively, given by
\begin{eqnarray}
\label{eq:bitobit}
|\Psi_1\rangle &=& 
(UB)^n|\alpha,\beta\rangle = \sum_{j_1,j_2,\cdots,j_n}
b_{j_n,j_{n-1}}\cdots b_{j_2,j_1}b_{j_1,\alpha}
|j_n,\beta+2J - n\rangle,\nonumber \\
|\Psi_2\rangle &=& 
(UB^{(1)})^n|\alpha,\beta\rangle = \sum_{j_1,j_2,\cdots,j_n}
b_{j_n,j_{n-1}}\cdots b_{j_2,j_1}b_{j_1,\alpha}
(e^{i\phi})^{j_{n-1}+ \cdots + j_1 + \alpha}
|j_n,\beta+2J - n\rangle,
\end{eqnarray}
where $J = j_1+\cdots+j_n$.  Consider an arbitrary state $|a,b\rangle$
in    the     computational-and-position    basis.    Now,    $\langle
a,b|\Psi_1\rangle  =  e^{i\eta\phi}\langle  a,b|\Psi_2\rangle$,  where
$\eta  = j_{n-1}+  \cdots +  j_1 +\alpha$,  which is  fixed  for given
$\alpha$ and $b$, and determined  by $b = \beta+2J-n$ and $j_n=a$.  As
a     result,     $|\langle     a,b|\Psi_1\rangle|^2    +     |\langle
\bar{a},b|\Psi_1\rangle|^2 =  |\langle a,b|\Psi_2\rangle|^2 + |\langle
\bar{a},b|\Psi_2\rangle|^2$.   A similar  proof of  invariance  of the
position distribution can be demonstrated to hold when $B$ is replaced
by  one of  the  other  $B^{(j)}$'s $(j=2,3,4)$.   On  account of  the
linearity of quantum mechanics,  the invariance of the walk statistics
under exchange of the $B^{(j)}$'s  and $B$ holds even when the initial
state $|\alpha,\beta\rangle$ is replaced by a general superposition or
a mixed state.  \hfill $\blacksquare$

Interchanging $B$ and the $B^{(j)}$'s may be considered as a discrete
symmetry operation $G: B \rightarrow B^{\star}$ (where $B^{\star}$
denotes any of the $B^{(j)}$'s in Eq. (\ref{eq:bias20})), 
that leaves the positional probability 
distribution invariant. We express this by the statement that
\begin{equation}
\label{eq:symm}
\widehat{W} \simeq {\bf G}\widehat{W},
\end{equation}
where ${\bf G}$  refers to the application of $G$ at  each step of the
walk, and $\widehat{W}$  refers to the walk operation  of evolving the
initial state  through $n$  steps and then  measuring in  the position
basis.  Knowledge of this symmetry can help simplify practical quantum
walks. Below we identify two of these quantum walk symmetry operations
$B   \leftrightarrow  B^{(1)}$   and   $B  \leftrightarrow   B^{(2)}$,
associated with physical operations of interest.

We   first  consider   the   phase  shift   gate  $\Phi(\phi)   \equiv
|0\rangle\langle0| +  e^{i\phi}|1\rangle\langle1|$ as a  {\em symmetry
operation} of a quantum walk.  In our model, the quantum operation for
each step is augmented by the insertion of $\Phi(\phi)$ just after the
operation $UB$.  At each step, the walker evolves according to:
\begin{eqnarray}
\label{eq:uz}
U_{\Phi} &\equiv& \left(\begin{array}{ll}
1 & 0 \\
0 & e^{i\phi}
\end{array}\right)
\left[
\left(\begin{array}{lcl}
1 & ~~ &0 \\
0 &  &0
\end{array}\right)\otimes \hat{A} +
\left(\begin{array}{lcl}
0 & & 0 \\
0 & &  1
\end{array}\right)\otimes \hat{A}^{\dag}\right]
\left( \begin{array}{clcr}
e^{i\xi}\cos(\theta)  & ~~ &  e^{i\zeta}\sin(\theta)   \\
e^{-i\zeta}\sin(\theta)  & &  -e^{-i\xi}\cos(\theta)
 \end{array} \right) \nonumber \\
&=&
\left[
\left(\begin{array}{lcl}
1 & ~~ &0 \\
0 &  &0
\end{array}\right)\otimes \hat{A} +
\left(\begin{array}{lcl}
0 & & 0 \\
0 & &  1
\end{array}\right)\otimes \hat{A}^{\dag}\right]
\left( \begin{array}{clcr}
e^{i\xi}\cos(\theta)  & ~~ &  e^{i\zeta}\sin(\theta)   \\
e^{i(\phi-\zeta)}\sin(\theta)  & &  -e^{i(\phi-\xi)}\cos(\theta)
 \end{array} \right).
\end{eqnarray}
This is equivalent to replacing  $B$ by $B^{(1)}$, which, according to
Theorem \ref{thm:bias}, leaves the walker distribution invariant. Thus
the operation $\Phi(\phi)$, applied at each step, is a symmetry of the
quantum walk.

As  a special case,  the phase  flip operation  ($Z$) applied  at each
step, obtained  by setting  $\phi=\pi$, is a  symmetry of  the quantum
walk.  Representing the inclusion of  operations $\Phi$ or $Z$ at each
step  of the  walk  by ${\bf  \Phi}$  or ${\bf  Z}$, respectively,  we
express this symmetry by the statements:
\begin{subequations}
\label{eq:sym1}
\begin{eqnarray}
\widehat{W} \simeq {\bf \Phi}\widehat{W}, \label{eq:sym1a} \\
\widehat{W} \simeq {\bf Z}\widehat{W}. \label{eq:sym1b} 
\end{eqnarray}
\end{subequations}

Unlike the  phase flip operation,  bit flip is  not a symmetry  of the
quantum walk on a line.  However, the combined application of bit flip
along with angular  reflection $R$ ($\theta \rightarrow \pi/2-\theta$,
i.e.,    $\sin\theta    \leftrightarrow    \cos\theta$,    and    $\xi
\leftrightarrow  -\zeta$)  and  parity $P$  ($\hat{A}  \leftrightarrow
\hat{A}^{\dag}$) turns  out to be  a symmetry operation.   These three
operations commute  with each other.   By the inclusion of  $PRX$, the
walker evolves  by $([PRX]  UB)^n$. At each  step, the  walker evolves
according to:
\begin{eqnarray}
U_P &\equiv& P R \left(\begin{array}{ll}
0 & 1 \\
1 & 0
\end{array}\right)
\left[
\left(\begin{array}{lcl}
1 & ~~ &0 \\
0 &  &0
\end{array}\right)\otimes \hat{A} +
\left(\begin{array}{lcl}
0 & & 0 \\
0 & &  1
\end{array}\right)\otimes \hat{A}^{\dag}\right]
\left( \begin{array}{clcr}
e^{i\xi}\cos(\theta)  & ~~ &  e^{i\zeta}\sin(\theta)   \\
e^{-i\zeta}\sin(\theta)  & &  -e^{-i\xi}\cos(\theta)
 \end{array} \right) \nonumber \\
&=&
\left[
\left(\begin{array}{lcl}
1 & ~~ &0 \\
0 &  &0
\end{array}\right)\otimes \hat{A} +
\left(\begin{array}{lcl}
0 & & 0 \\
0 & &  1
\end{array}\right)\otimes \hat{A}^{\dag}\right]
\left( \begin{array}{clcr}
e^{i\xi}\cos(\theta)  & ~~ &  -e^{i\zeta}\sin(\theta)   \\
e^{-i\zeta}\sin(\theta)  & &  e^{-i\xi}\cos(\theta)
 \end{array} \right).
\end{eqnarray}
This  is equivalent  to replacing  $B$ by  $B^{(2)}$  with $\phi=\pi$,
which, according  to Theorem  \ref{thm:bias}, should leave  the walker
distribution invariant. Thus the operation $PRX$ applied at each step,
is a symmetry  of the quantum walk.  It  will be convenient henceforth
to choose  $\xi=\zeta=0$, so  that $R$ will  simply correspond  to the
replacement $\theta \rightarrow \pi/2-\theta$.

Representing the inclusion of operations  $P$, $R$ or $X$ at each step
of the  walk by ${\bf  P}$, ${\bf R}$  or ${\bf X}$,  respectively, we
express this symmetry by the statements:
\begin{subequations}
\label{eq:RX}
\begin{eqnarray}
\widehat{W} &\simeq& {\bf P R X}\widehat{W}, \label{eq:RX1} \\
{\bf X}\widehat{W} &\simeq& {\bf P R} \widehat{W}. \label{eq:RX2}
\end{eqnarray}
\end{subequations}
Eq.  (\ref{eq:RX1}) was proved  immediately above.  Eq. (\ref{eq:RX2})
follows from Eq. (\ref{eq:RX1}), since the operations ${\bf P}$, ${\bf
R}$  and ${\bf  X}$  mutually  commute, and  $X^2  = \mathbb{I}$.   It
expresses that  fact that applying the  $X$ operation at  each step is
equivalent  to  replacing  a  quantum  walk  by  its  angle-reflected,
spatially inverted counterpart. The  observation made at the beginning
of this  Section pertains to the special  case of $\theta=45^{\circ}$.
By a similar technique the following symmetries may be proved:
\begin{equation}
\label{eq:sym3}
{\bf X}\widehat{W} ~\simeq~ {\bf X Z}\widehat{W}
~\simeq~ {\bf Z X}\widehat{W}.
\end{equation}
The first equivalence easily follows from
Eq. (\ref{eq:sym1}).

\section{Environmental effects \label{sec:env}}
\begin{figure}
\subfigure[]{\includegraphics[width=8.6cm]{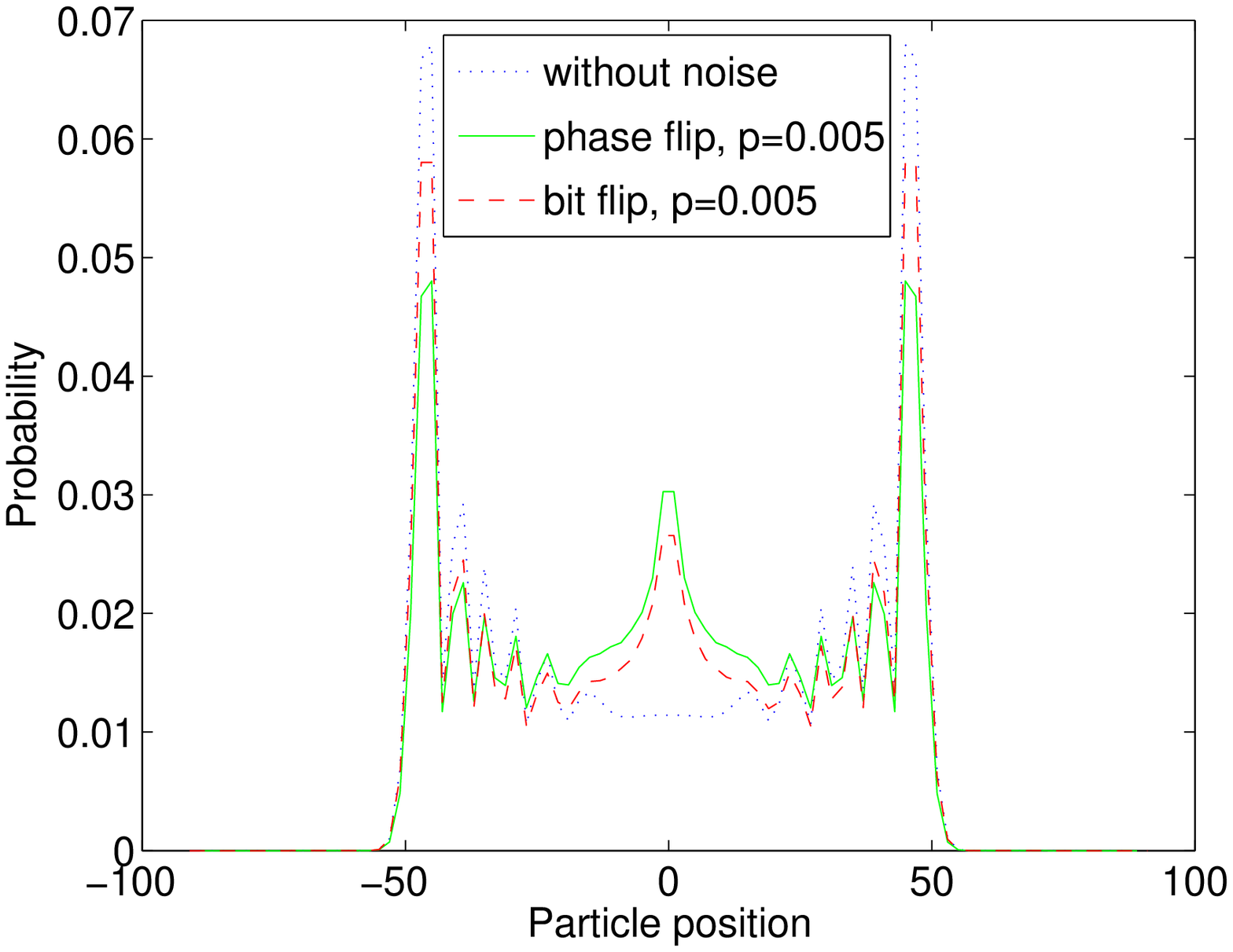}}
\hfill
\subfigure[]{\includegraphics[width=8.6cm]{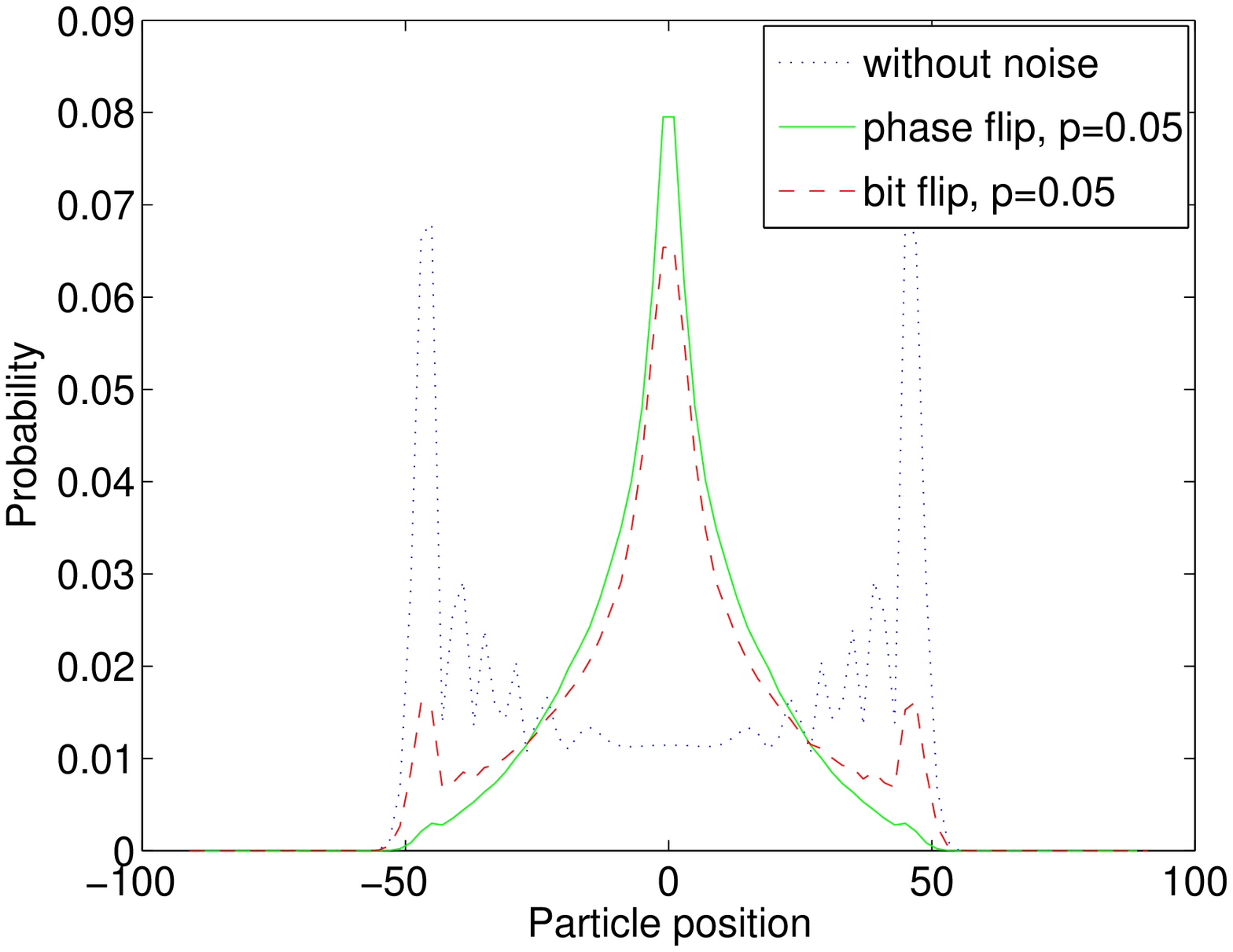}} \\
\subfigure[]{\includegraphics[width=8.6cm]{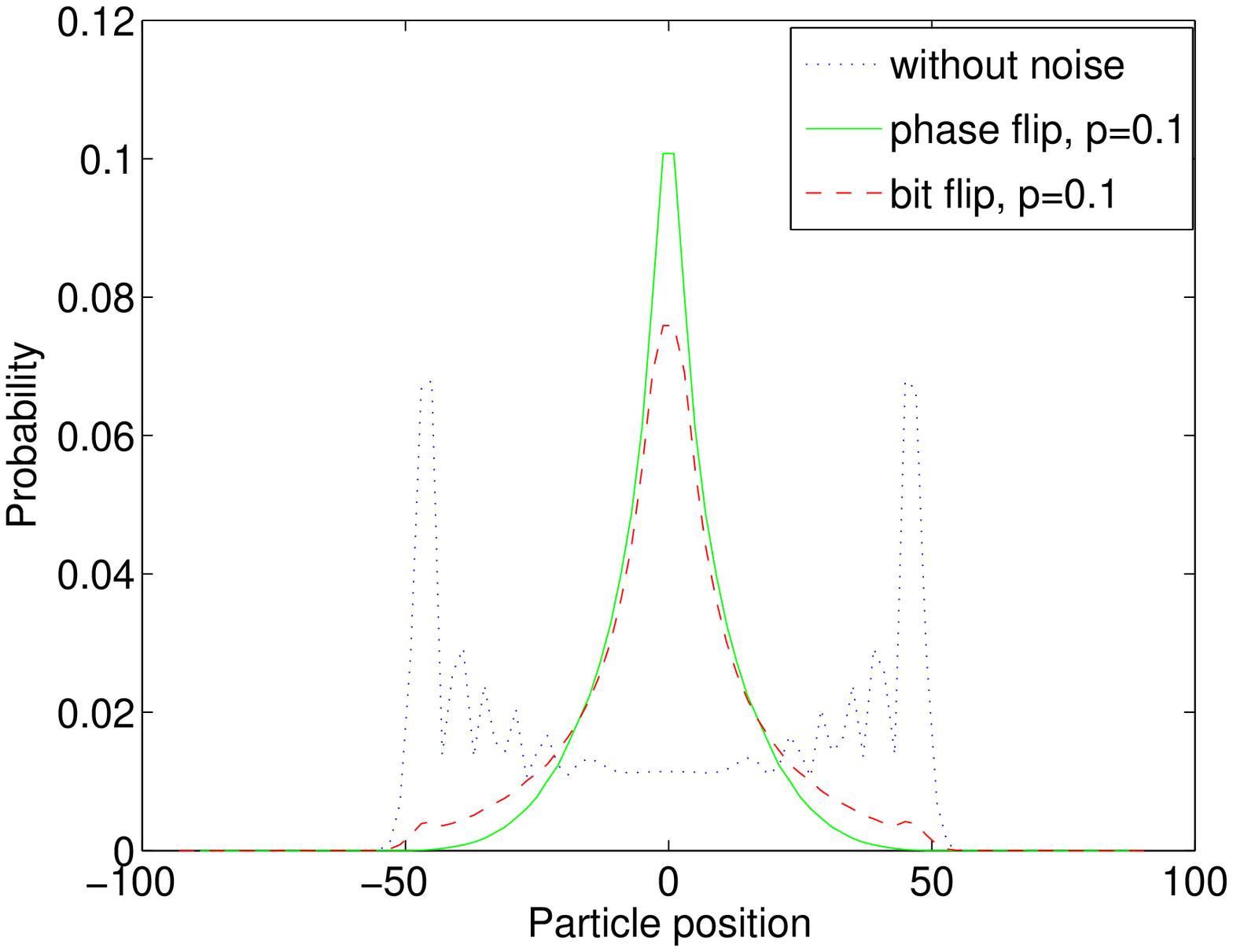}}
\hfill
\subfigure[]{\includegraphics[width=8.6cm]{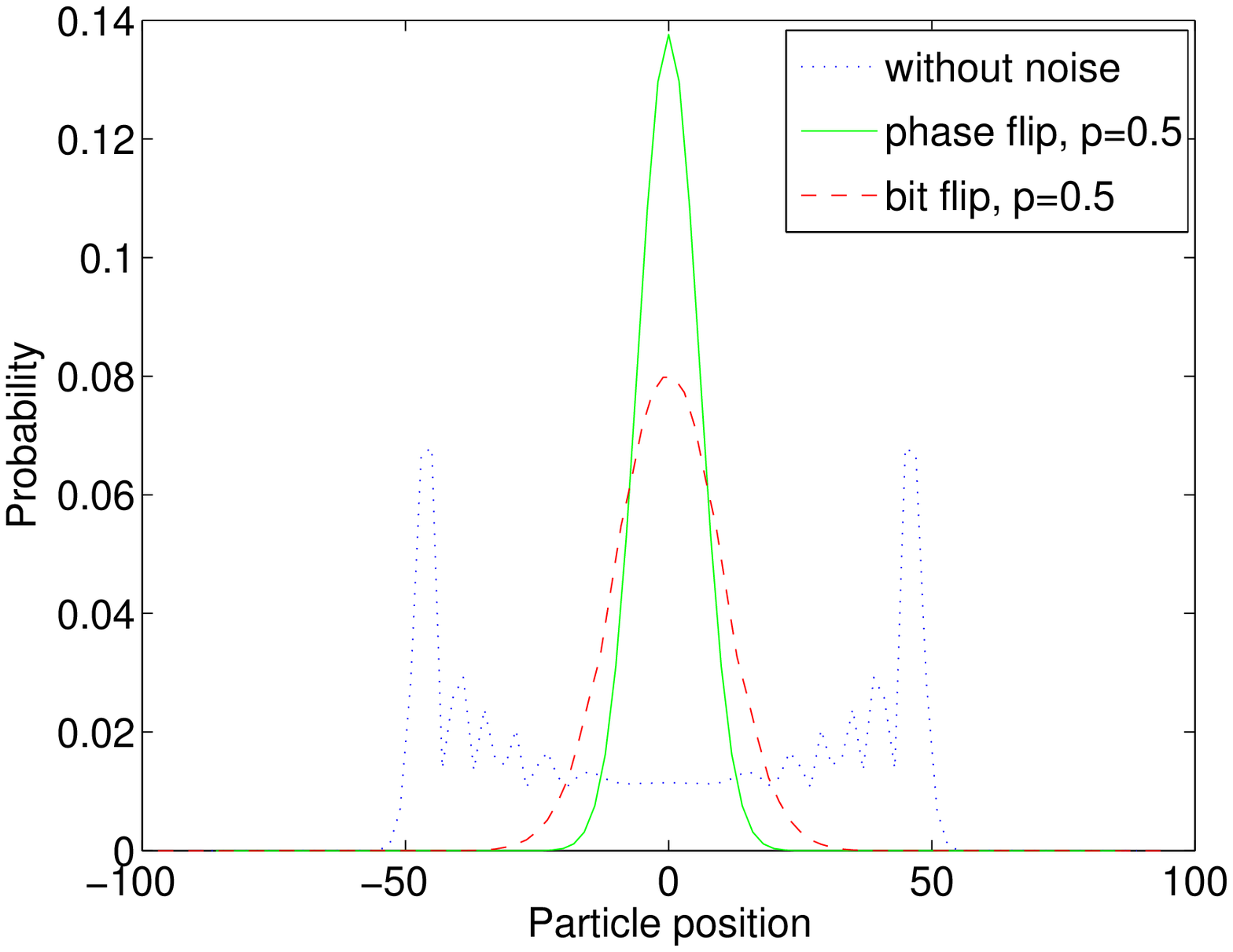}}
\caption{ (color online) The  effect  of  environmental  decoherence on  the  position
probability  distribution of a  biased quantum  walker subjected  to a
noisy  channel. Coin  bias is  of the  form Eq.  (\ref{eq:bias1}) with
$\theta = 60^{\circ}$.   The noise is modelled as  a phase flip (solid
line)   or  bit   flip   (dashed  line)   channel,  characterized   by
Eqs.  (\ref{eq:phaseflip})  and  (\ref{eq:bitflip}), respectively,  at
various noise levels  $p$: (a) $p = 0.005$  (b) $p=0.05$; (c) $p=0.1$;
(d)   $p=0.5$,  which   corresponds  to   a  fully   classical  random
walk.  Comparing Figure  (d) with  Figure \ref{fig:env30}(d),  we note
that the  distribution in the  case of maximal  bit flip noise  is the
same.}
\label{fig:env60}
\end{figure}

\begin{figure}
\subfigure[]{\includegraphics[width=8.6cm]{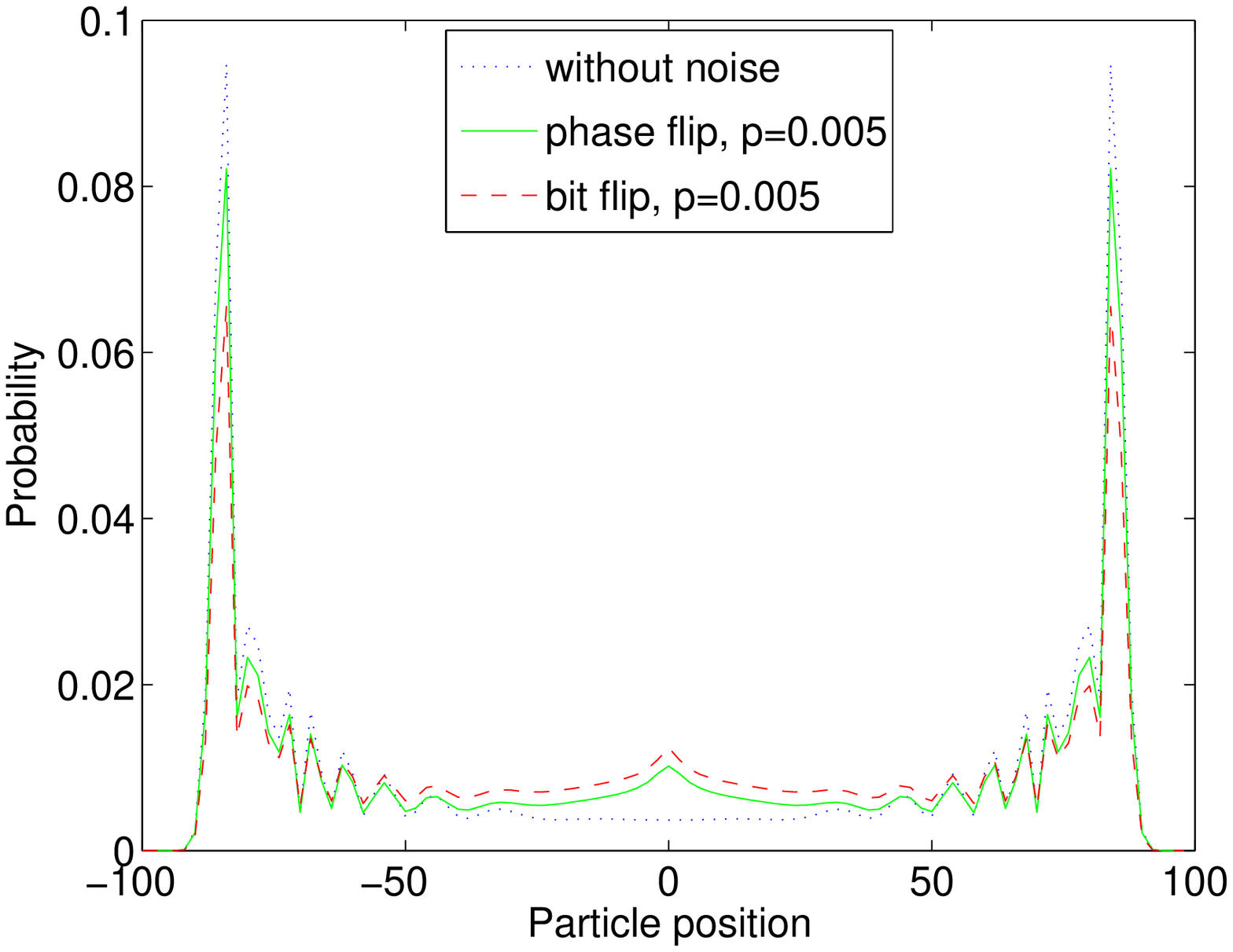}}
\hfill
\subfigure[]{\includegraphics[width=8.6cm]{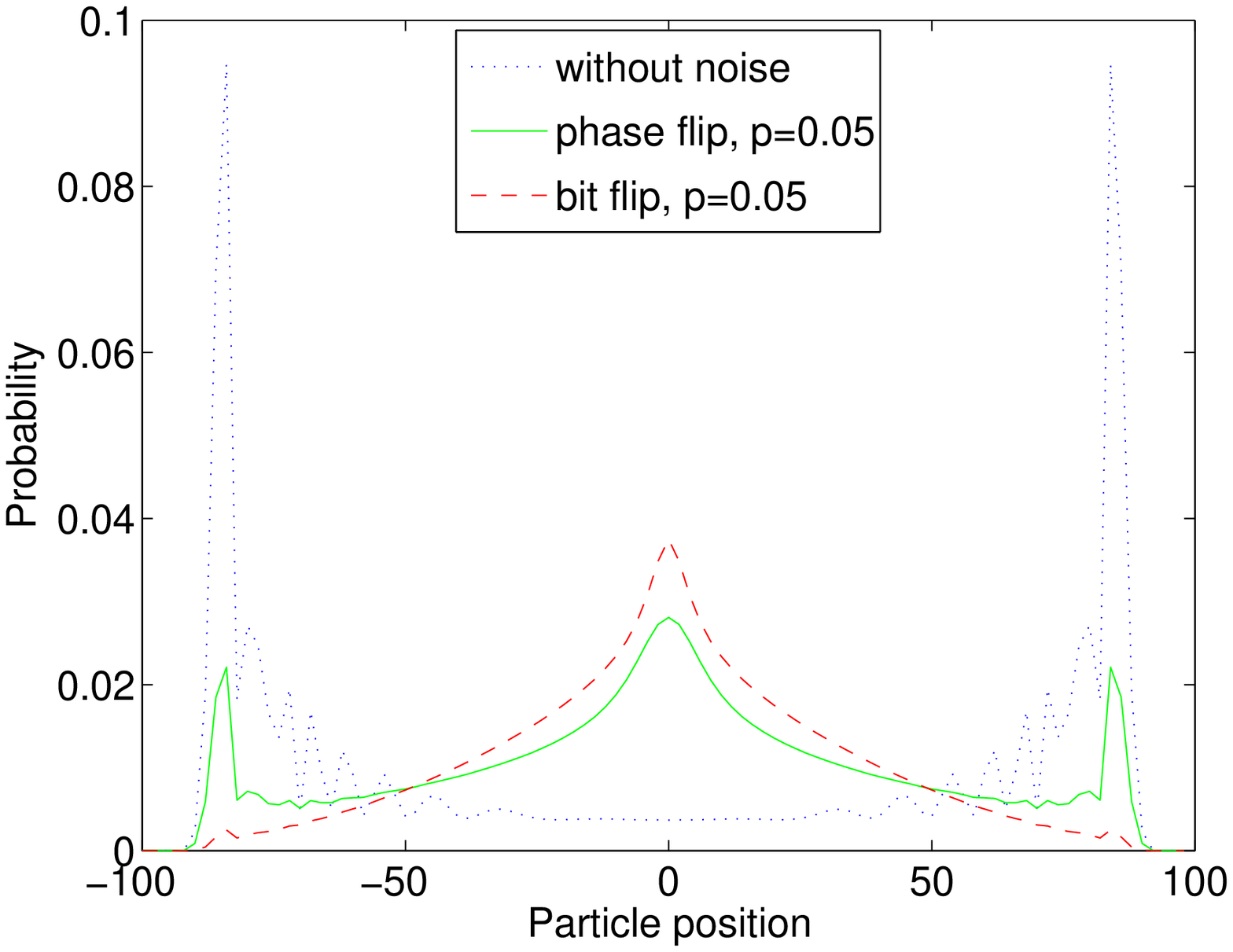}} \\
\subfigure[]{\includegraphics[width=8.6cm]{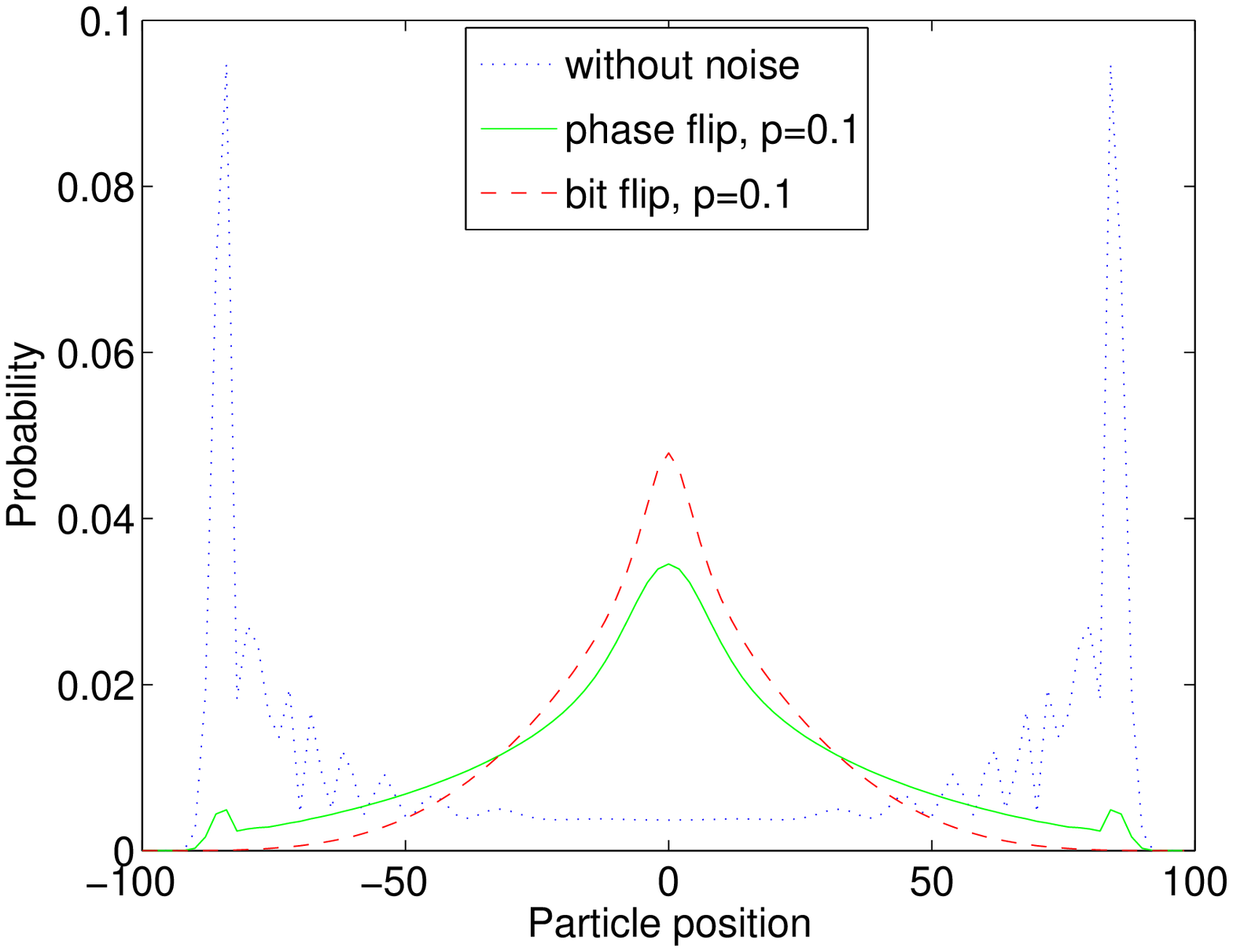}}
\hfill
\subfigure[]{\includegraphics[width=8.6cm]{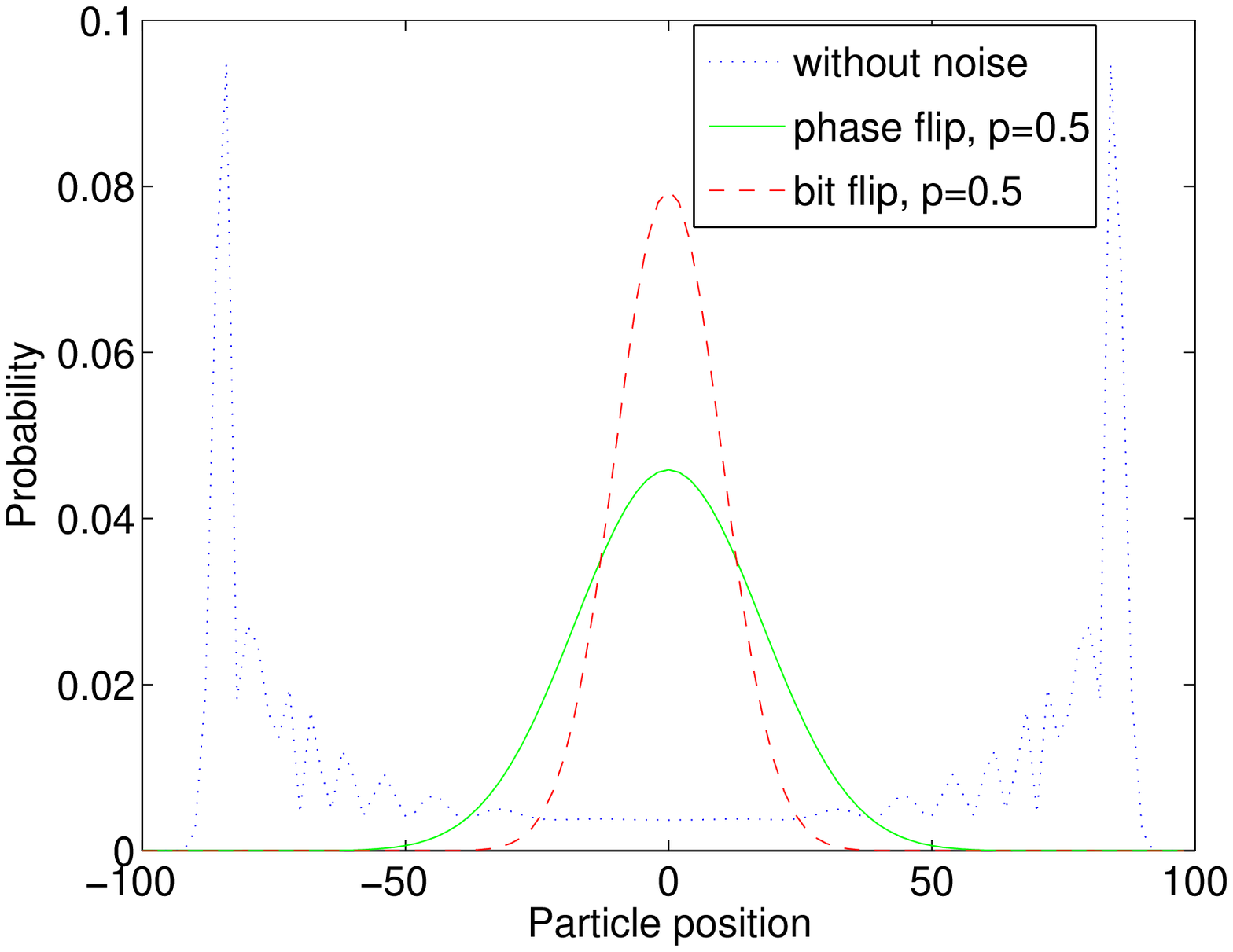}}
\caption{(color online) The  effect  of  environmental  decoherence on  the  position
probability  distribution of a  biased quantum  walker subjected  to a
noisy  channel. Coin  bias is  of the  form Eq.  (\ref{eq:bias1}) with
$\theta = 30^{\circ}$.   The noise is modelled as  a phase flip (solid
line)   or  bit   flip   (dashed  line)   channel,  characterized   by
Eqs.  (\ref{eq:phaseflip})  and  (\ref{eq:bitflip}), respectively,  at
various noise levels  $p$: (a) $p = 0.005$  (b) $p=0.05$; (c) $p=0.1$;
(d) $p=0.5$, which corresponds to a fully classical random walk.}
\label{fig:env30}
\end{figure}

\begin{figure}
\includegraphics[width=8.6cm]{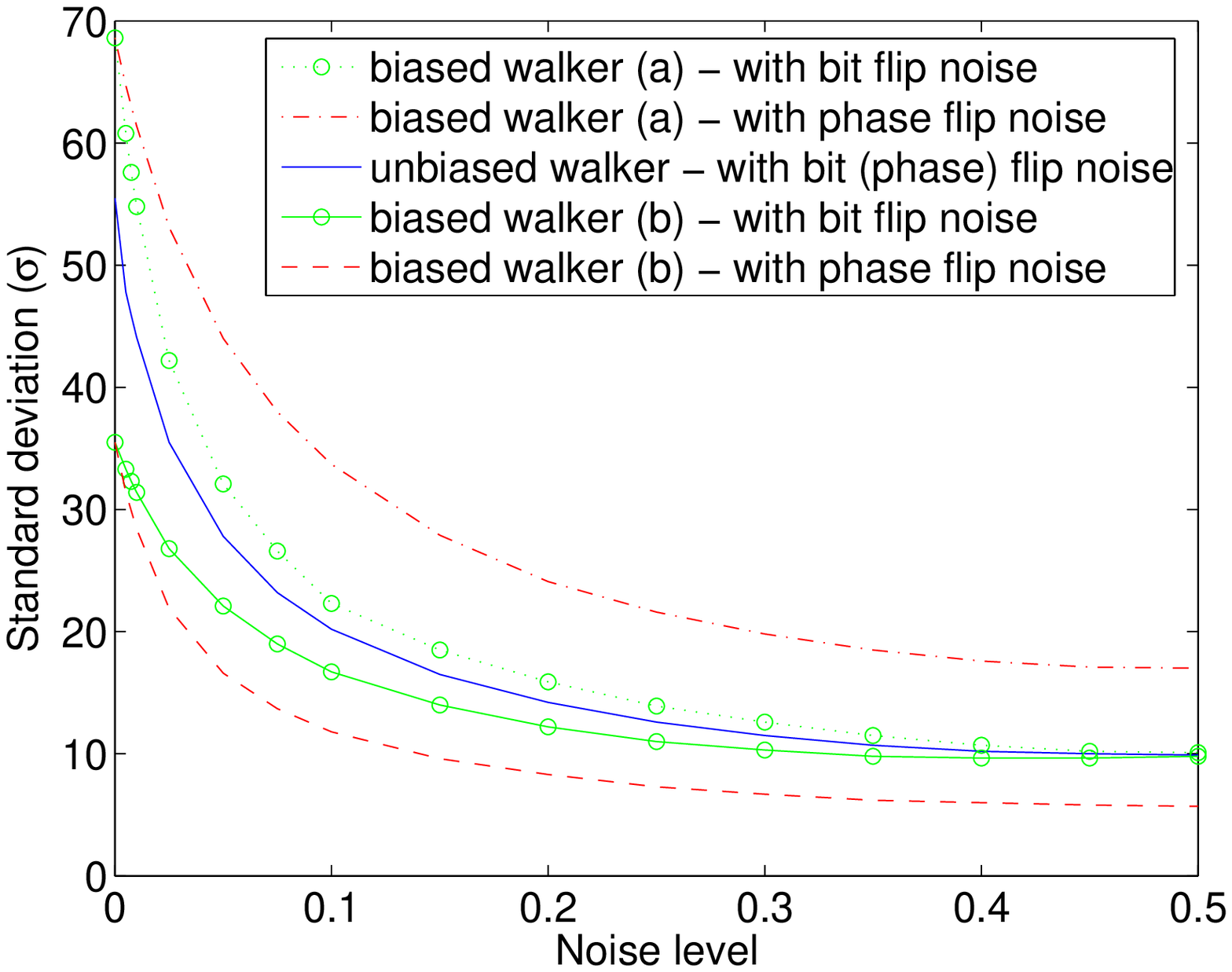}
\caption{(color online) Variation of  standard deviation  with noise level,  for both
phase  noise  and  bit flip  noise.  Solid  line  is for  an  unbiased
walker.  For biased walker  (a) $\theta  =30^{\circ}$ and  (b) $\theta
=60^{\circ}$.   In  the  classical  limit  of  $p=0.5$,  the  standard
deviation converges to a fixed  value for bit flip noise, irrespective
of  $\theta$, but  different  for phase  flip  noise. The  convergence
happens because, at maximum  bit flip noise ($p=0.5$), the measurement
outcome in  the computational basis is completely  randomized, and the
presence  or absence of  bias is  irrelevant. On  the other  hand, the
non-convergence in  the case of  phase flip noise  is due to  the fact
that the asymptotic  mixed state obtained via a  phase damping channel
depends on the initial state parameter $\theta$.}
\label{fig:sd}
\end{figure}

A  quantum  walk  implemented  on  a quantum  computer  is  inevitably
affected by errors caused by noise due to the environment. We consider
three physically relevant models of noise: a phase flip channel (which
is equivalent to  a phase damping or purely  dephasing channel), a bit
flip channel and a generalized  amplitude damping channel ($T \ge 0$).
In all  cases, our numerical implementation of  these channels evolves
the  density matrix  employing the  Kraus operator  representation for
them. However to explain symmetry effects, it is convenient to use the
{\em quantum trajectories} approach, discussed below.

\subsection{Decoherence via phase damping and bit flip 
channels \label{sec:enva}}

In  studying the  status of  the walk  symmetries in  the  presence of
noise, it is advantageous  to employ the quantum trajectories approach
\cite{bru02}.  This  simplifies the  description  of  an open  quantum
system in terms of a  stochastically evolving pure state, which allows
us to  adapt the symmetry  results for the  pure states, given  in the
preceding section, to mixed states.

We  call the  sequence  of walk  step  operations $\widehat{X}  \equiv
(UB_n)(UB_{n-1})\cdots(UB_1)$    a   `quantum    trajectory'.    (More
precisely, a trajectory  refers to the sequence of  states produced by
these  operations,  for  which   the  above  serves  as  a  convenient
representation.)  If all the  $B_j$'s are the same, then $\widehat{X}$
is the usual `homogeneous' quantum walk $\widehat{W}$. In general, the
$B_j$'s may be different $SU(2)$ operators, corresponding to a varying
bias in the coin degree of  freedom.  More generally, each step of the
walk  may include  generalized measurements  whose outcomes  are known
(Section  \ref{sec:envb}).   If each  walk  step  in $\widehat{X}$  is
subjected  to a  fixed symmetry  operation $G$,  the result  is  a new
quantum       trajectory        ${\bf       G}\widehat{X}       \equiv
(UB_n^{\star})(UB_{n-1}^{\star})\cdots  (UB_1^{\star})$.  We  have the
following  generalization of  Theorem \ref{thm:bias}  to inhomogeneous
quantum walks on a line.

\begin{thm}
Given any quantum  walk trajectory $\widehat{X} = (UB_n)\cdots(UB_1)$,
the  symmetry  ${\cal  G}$   holds,  i.e.,  $\widehat{X}  \simeq  {\bf
G}\widehat{X}$. If  the operation ${\bf \Phi}$ is  restricted to ${\bf
Z}$, then the symmetries hold even when some of the $U$'s are replaced
by $U^{\dag}$'s.
\label{thm:qutraj}
\end{thm}
{\bf Proof.}  In the proof of Theorem \ref{thm:bias}, we note that if,
in  each  step  of  the  walk,  we  alter  the  rotation  $B$  by  the
transformation  $G$,   the  proof   still  goes  through.    That  is,
$|\Psi_1\rangle                                                  \equiv
(UB_n)(UB_{n-1})\cdots(UB_1)|\alpha,\beta\rangle$  and $|\Psi_2\rangle
\equiv
(UB_n^{\star})(UB_{n-1}^{\star})\cdots(UB_1^{\star})|\alpha,\beta\rangle$
produce the same position distribution.

Suppose  that  in  some  of  the  walk steps,  $U$  is  replaced  with
$U^{\dag}$.  In place of Eq. (\ref{eq:bitobit}), we have
\begin{subequations}
\label{eq:bitobitb}
\begin{eqnarray}
\label{eq:bitobitb1}
|\Psi_1\rangle &=& 
(UB_n)\cdots(U^{\dag}B_{j})\cdots(UB_1)|\alpha,\beta\rangle\nonumber\\
&=& \sum_{j_1,j_2,\cdots,j_n}
b_{j_n,j_{n-1}}\cdots b_{j_2,j_1}b_{j_1,\alpha}
|j_n,\beta+2(J_1-J_2) - (n_1-n_2)\rangle,\\
\label{eq:bitobitb2}
|\Psi_2\rangle &=& 
(UB_n^{(1)})\cdots(U^{\dag}B_{j}^{(1)})\cdots(UB_1^{(1)})|\alpha,\beta\rangle
\nonumber\\
&=& \sum_{j_1,j_2,\cdots,j_n}
b_{j_n,j_{n-1}}\cdots b_{j_2,j_1}b_{j_1,\alpha}
(-1)^{j_{n-1}+ \cdots + j_1 + \alpha}
|j_n,\beta+2(J_1-J_2) - (n_1-n_2)\rangle,
\end{eqnarray}
\end{subequations}
where $J_1 = \sum_k j_k$ for the $n_1$ steps $k$ where operator $U$ is
used, and  $J_2 = \sum_l j_l$  for the $n_2$ steps  $l$ where operator
$U^{\dag}$ is  used.  Here $J=J_1+J_2$ and  $n=n_1+n_2$.  Observe that
the  exponent   of  $(-1)$   is  effectively  evaluated   in  modulo-2
arithmetic. We  can thus  replace $J$ by  $J_1 -J_2$ in  the exponent.
Following  the  argument  in  Theorem  \ref{thm:bias},  we  find  that
$\langle  a,b|\Psi_1\rangle = e^{i\Theta}  \langle a,b|\Psi_2\rangle$,
where $\Theta = J_1-J_2-a+\alpha$.  \hfill $\blacksquare$
\bigskip

As a  corollory, the symmetries  ${\bf Z}$ and  ${\bf P R  X}$ hold
good because they reduce to special cases of ${\bf G}$.
A question of practical interest is whether
${\bf PRX}$ and ${\bf Z}$ are symmetries of a noisy quantum walk.
Suppose we are given a noise process ${\cal N}$ in the Kraus representation
\begin{equation}
\label{eq:kraus}
\rho \longrightarrow {\cal N}(\rho) = \sum_{j=0}^{m-1} E_j\rho E^{\dag}_j,
\hspace{0.5cm} \sum_j E^{\dag}_jE_j=\mathbb{I}.
\end{equation}
With the  inclusion of  noise, each step  of the quantum  walk becomes
augmented  to $(\Pi U  B_k)$, where  $\Pi$ is  a random  variable that
takes Kraus operator  values $E_j$. Thus, ${\cal N}$  corresponds to a
mixture  of upto  $n^m$ trajectories  or  `unravelings' $\widehat{X}_l
\equiv \left(\Pi(l_n)UB_n\right)\cdots\left(\Pi(l_1)UB_1\right)$, each
occuring with some probability  $p_l$, where $\sum_l p_l=1$.  If ${\bf
Z}$ and  ${\bf PRX}$ are symmetries of  an unraveling $\widehat{X}_l$,
then the  operations $\widehat{X}_l$ and  ${\bf H}\widehat{X}_l \equiv
\left(\Pi(l_n)HUB_n\right)\cdots\left(\Pi(l_1)HUB_1\right)$,      where
${\bf  H}$ denotes  ${\bf  Z}$ or  ${\bf  PRX}$, must  yield the  same
position  probability  distribution.   In  the case  of  bit-flip  and
phase-flip channels,  there is a  representation in which  the $E_j$'s
are proportional to unitary operators. Further:
\begin{thm}
If  trajectories  $\widehat{X}_l$  are  individually  symmetric  under
operation ${\bf G}$, then so  is any noisy quantum walk represented by
a collection $\{\widehat{X}_l, p_l\}$.
\label{thm:mix}
\end{thm}
{\bf Proof.}  The state  of the  system obtained via  ${\cal N}$  is a
linear  combination   (the  average)   of  states  obtained   via  the
$\widehat{X}_j$'s. Thus, the invariance of the $\widehat{X}_j$'s under
${\bf G}$ implies the invariance of the former.  \hfill $\blacksquare$
\bigskip

This result, together with those from the preceding Section, can
now be easily shown to imply that the symmetry ${\bf H}$ is preserved
in the case of phase-flip and bit-flip channels.

Decoherence via a purely dephasing channel, without any loss of
energy, can be modeled as a phase flip channel \cite{nc00,srib06}:
\begin{equation}
\label{eq:phaseflip}
{\cal E}(\rho) = (1-p)\rho + pZ\rho Z.
\end{equation}
An example of a physical process that realizes Eq. (\ref{eq:phaseflip}) 
is a two-level system interacting with its bath via
a quantum non-demolition (QND) interaction given by the Hamiltonian 
\begin{equation}
H =  H_S + \sum\limits_k \hbar \omega_k b^{\dagger}_k b_k + H_S 
\sum\limits_k g_k (b_k+b^{\dagger}_k) + H^2_S \sum\limits_k 
{\frac {g^2_k}{\hbar \omega_k}}. \label{2a} 
\end{equation} 
Here $H_S$ is the system Hamiltonian and 
the second term on the RHS of the above equation
is the free Hamiltonian of the environment, while the third
term is the system-reservoir interaction Hamiltonian.
The last term on the RHS of Eq. (\ref{2a})
is a renormalization inducing `counter term'. Since $[H_S, 
H_{SR}]=0$, Eq. (\ref{2a}) is of QND type.

Following Ref. \cite{srib06} (apart from a change in notation
which switches $|0\rangle \longleftrightarrow |1\rangle$),
taking into account the effect of the
environment modeled as a thermal bath, 
the reduced dynamics of the system 
can be obtained, which can be described using Bloch vectors
as follows. Its action on an initial state:
\begin{equation}
\rho_0 \equiv \begin{pmatrix}
\frac{1}{2}\left(1 + \langle \sigma_3(0) \rangle
\right) & \langle \sigma_-(0) \rangle \cr
\langle \sigma_+(0) \rangle & {\frac {1}{2}} \left(1 -
\langle \sigma_3(0) \rangle \right)
\end{pmatrix},
\label{eq:inista}
\end{equation}
is given in the interaction picture by:
\begin{equation}
\label{eq:qnd2}
{\cal E}(\rho_0) = \begin{pmatrix}
\frac{1}{2}\left(1 + \langle \sigma_3(0) \rangle
\right) & \langle \sigma_-(0) \rangle e^{-(\hbar\omega)^2\gamma(t)} \cr
\langle \sigma_+(0) \rangle e^{-(\hbar\omega)^2\gamma(t)} & 
{\frac{1}{2}} \left(1 -
\langle \sigma_3(0) \rangle \right)
\end{pmatrix}.
\end{equation}
The initial state (\ref{eq:inista}) may be mixed. (The derivation 
of the superopertor ${\cal E}$ in terms of environmental parameters for
the pure state case, given explicitly in Ref. \cite{srib06}, 
is directly generalized to the case of an arbitrary mixture of
pure states, since the environmental parameters are assumed to be
independent of the system's state.)

Comparing Eq. (\ref{eq:qnd2}) with Eq. (\ref{eq:phaseflip}) allows us
to relate the noise level $p$ in terms of physical parameters.
In particular:
\begin{equation}
p = \frac{1}{2}\left(1 - \exp\left[-(\hbar\omega)^2\gamma(t)\right]\right).
\end{equation}
When $\gamma(t)\approx0$ (either because the coupling with the environment
is very weak or the interaction time is short or the temperature is low),
$p\approx0$, tending towards the noiseless case. On the other hand,
under strong coupling, $\gamma(t)$ is arbitrarily large, and $p\rightarrow1/2$,
the maximally noisy limit.
The result of implementing channel (\ref{eq:phaseflip}) is to
drive the position
probability distribution towards a classical Gaussian
pattern \cite{ken05}. The effect of increasing phase noise in the
presence of biased walk is depicted
in Figure \ref{fig:env60}, for the case of $\theta=60^{\circ}$,
and in Figure \ref{fig:env30}, for the case of $\theta=30^{\circ}$.
The onset of classicality is observed in the Gaussianization
of the probability distribution. This is reflected also in the fall of
standard deviation, as shown in Figure \ref{fig:sd}.

Decoherence can also be introduced by another noise model,
the bit flip channel \cite{nc00}:
\begin{equation}
\label{eq:bitflip}
{\cal E}(\rho) = (1-p)\rho + pX\rho X.
\end{equation}
As with the phase damping channel,
the bit flip channel also
drives the probability distribution towards a classical, Gaussian
pattern, with increasing noise \cite{ken05}. 
The effect of increasing bit flip noise in the
presence of biased walk is depicted
in Figures \ref{fig:env60} and \ref{fig:env30}.
Here again, the onset of classicality is observed in the Gaussianization
of the probability distribution, as well as in the fall of
standard deviation, as shown in Figure \ref{fig:sd}.

A difference in  the classical limit of these  two noise processes, as
observed  in  Figure  \ref{fig:sd},   is  that  whereas  the  standard
deviation (in fact, the distribution) is unique in the case of the bit
flip channel  irrespective of bias, in  the case of  phase flip noise,
the classical  limit distribution is  bias dependent. This  is because
phase flip  noise leads,  in the Bloch  sphere picture, to  a coplanar
evolution of states towards the $\sigma_z$ axis. Thus all initial pure
states corresponding to a  fixed $\theta$ evolve asymptotically to the
same   mixed  state   \cite{nc00,srib06}.  This   also   explains  the
contrasting behavior of bit flip  and phase flip noise with respect to
bias,   as   seen   by   comparing   Figures   (\ref{fig:env60})   and
(\ref{fig:env30}).

\begin{figure}
\includegraphics[width=8.6cm]{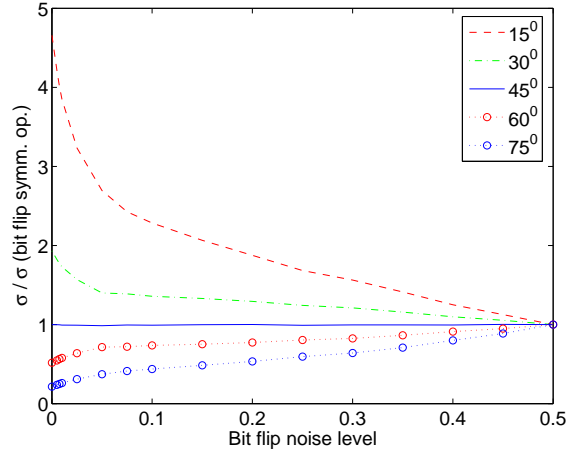}
\caption{(color online) Variation  of the  ratio  of standard  deviation without  any
symmetry operation to the  bit flip symmetry operation with increasing
bit flip noise level.}
\label{fig:sd2}
\end{figure}

\begin{figure}
\includegraphics[width=8.6cm]{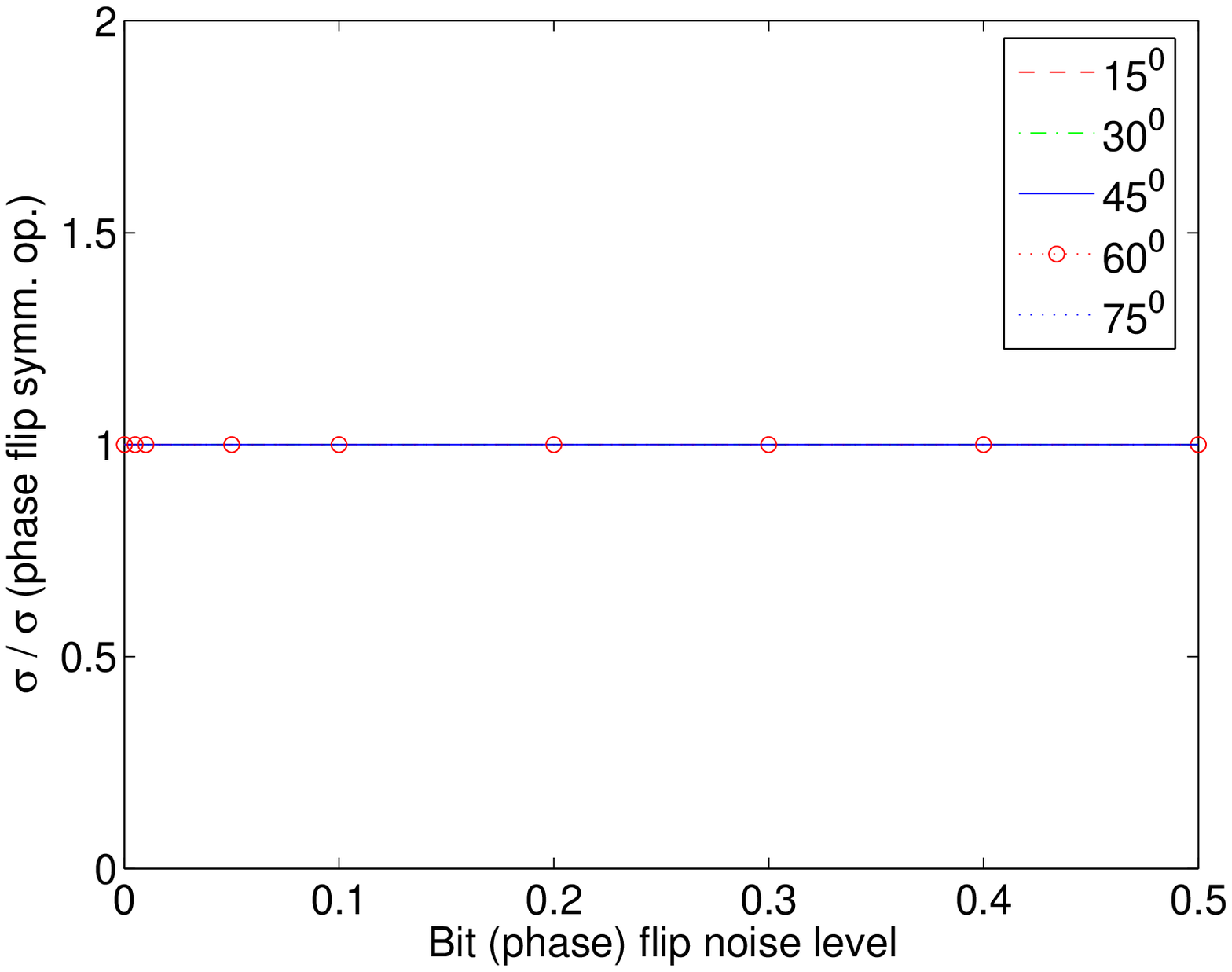}
\caption{(color online) Variation  of the  ratio  of standard  deviation without  any
symmetry  operation   to  the  phase  flip   symmetry  operation  with
increasing bit flip (phase flip) noise level.}
\label{fig:sd3}
\end{figure}

\begin{figure}
\includegraphics[width=8.6cm]{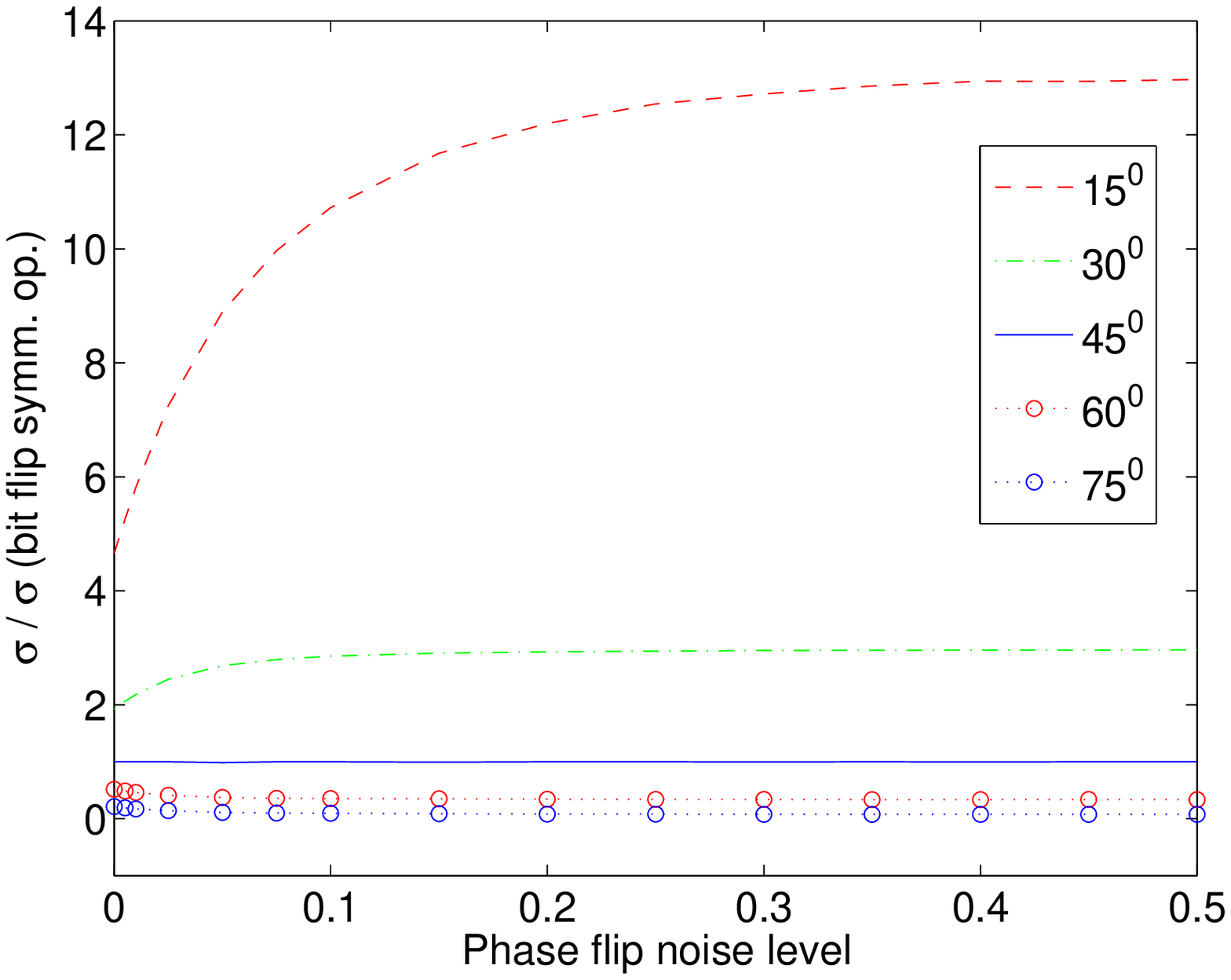}
\caption{(color online)  Variation  of the  ratio  of standard  deviation without  any
symmetry operation to the bit symmetry operation with increasing phase
flip noise level.}
\label{fig:sd5}
\end{figure}

\begin{figure}
\subfigure[]{\includegraphics[width=8.6cm]{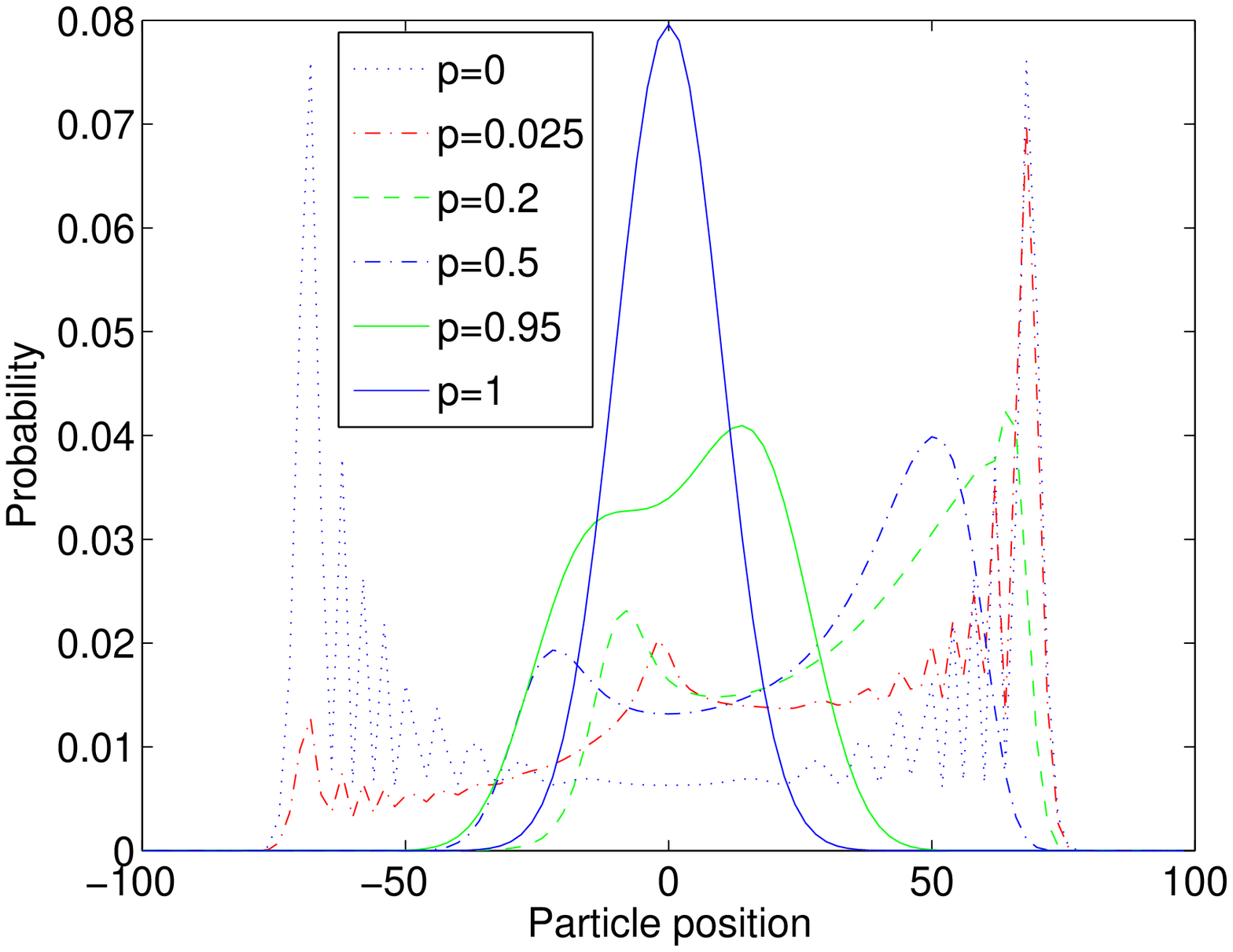}}
\subfigure[]{\includegraphics[width = 8.6cm]{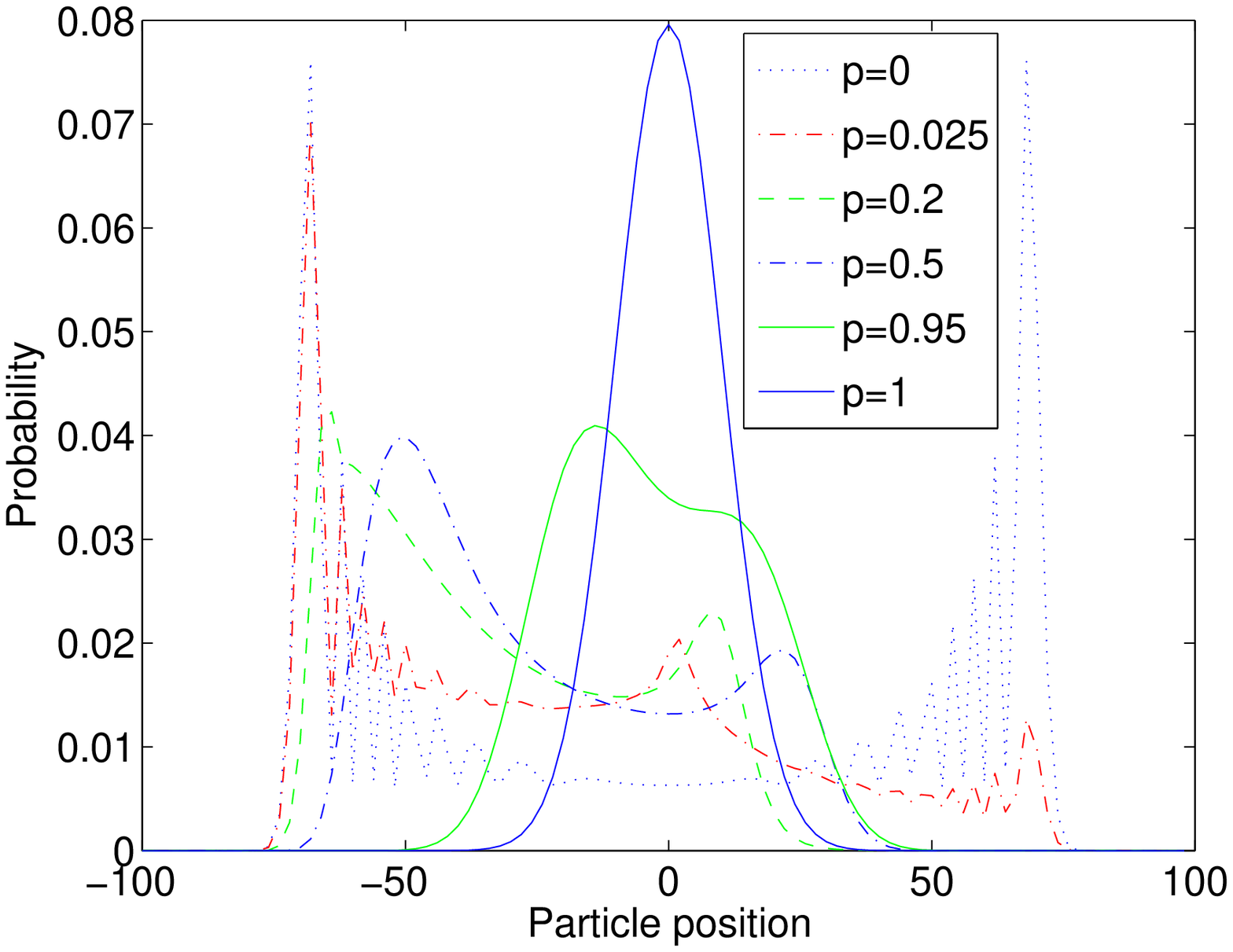}}
\caption{(color online) Amplitude damping channel acting on a Hadamard walker 
at temperature $T=0$. The distribution corresponding to intermediate 
values of $p$ clearly show the breakdown of the ${\bf R X}$ symmetry.
However, the extended symmetry, ${\bf P R X}$ (where ${\bf P}$ stands
for parity operation (spatial inversion)) holds good. This is observed
at all temperatures.
(a) Probability distribution of finding the particle 
undergoing unbiased quantum walk on which
amplitude damping  channel is acting.  This shows that
even at $T=0$, for sufficiently high coupling, the distribution turns
classical. (b) Amplitude damping  with bit
flip symmetry. }
\label{fig:ampdamp}
\end{figure}

\begin{figure}
\subfigure[]{\includegraphics[width=8.6cm]{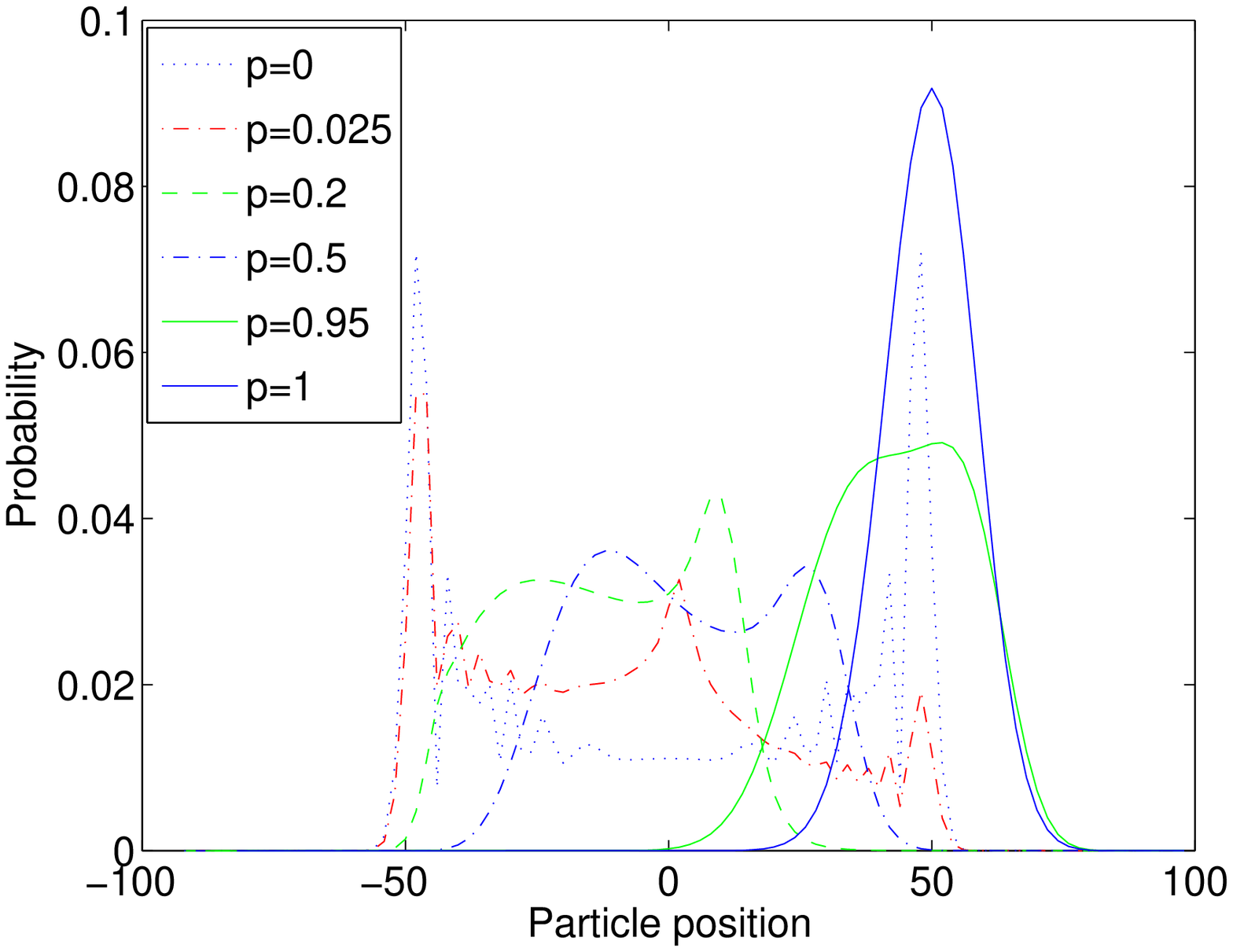}}
\hfill
\subfigure[]{\includegraphics[width=8.6cm]{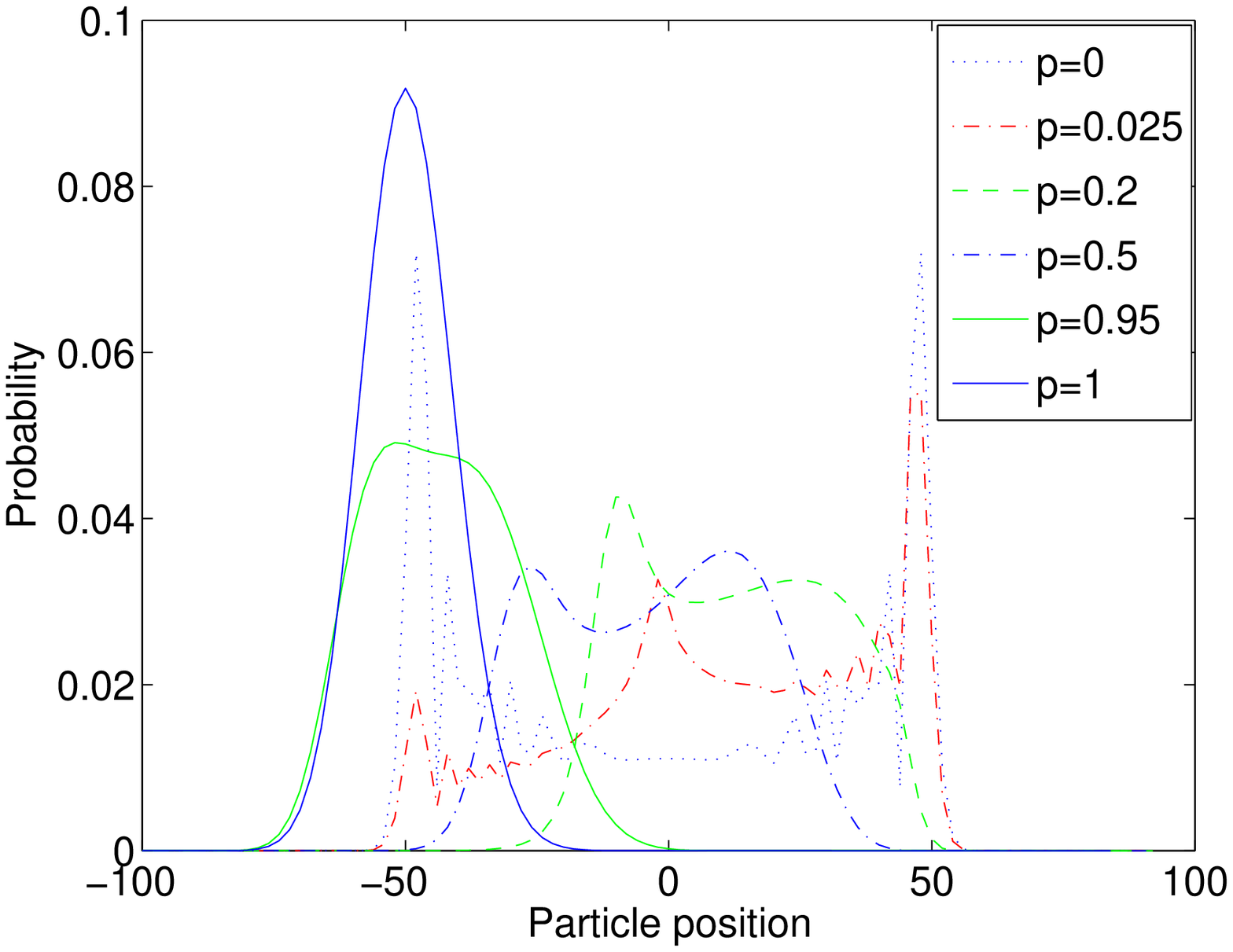}}
\caption{(color online) ${\bf PRX}$ symmetry seen to hold in biased quantum walk
subjected to amplitude damping ($T=0$). The two cases are spatial inversions
of each other. This holds for a generalized amplitude damping
at any temperature. (a) walker with $\theta =
30^{\circ}$ and bit flip; (b) $\theta = 60^{\circ}$.}
\label{fig:ampdamp34}
\end{figure}

\begin{figure}
\subfigure[]{\includegraphics[width=8.6cm]{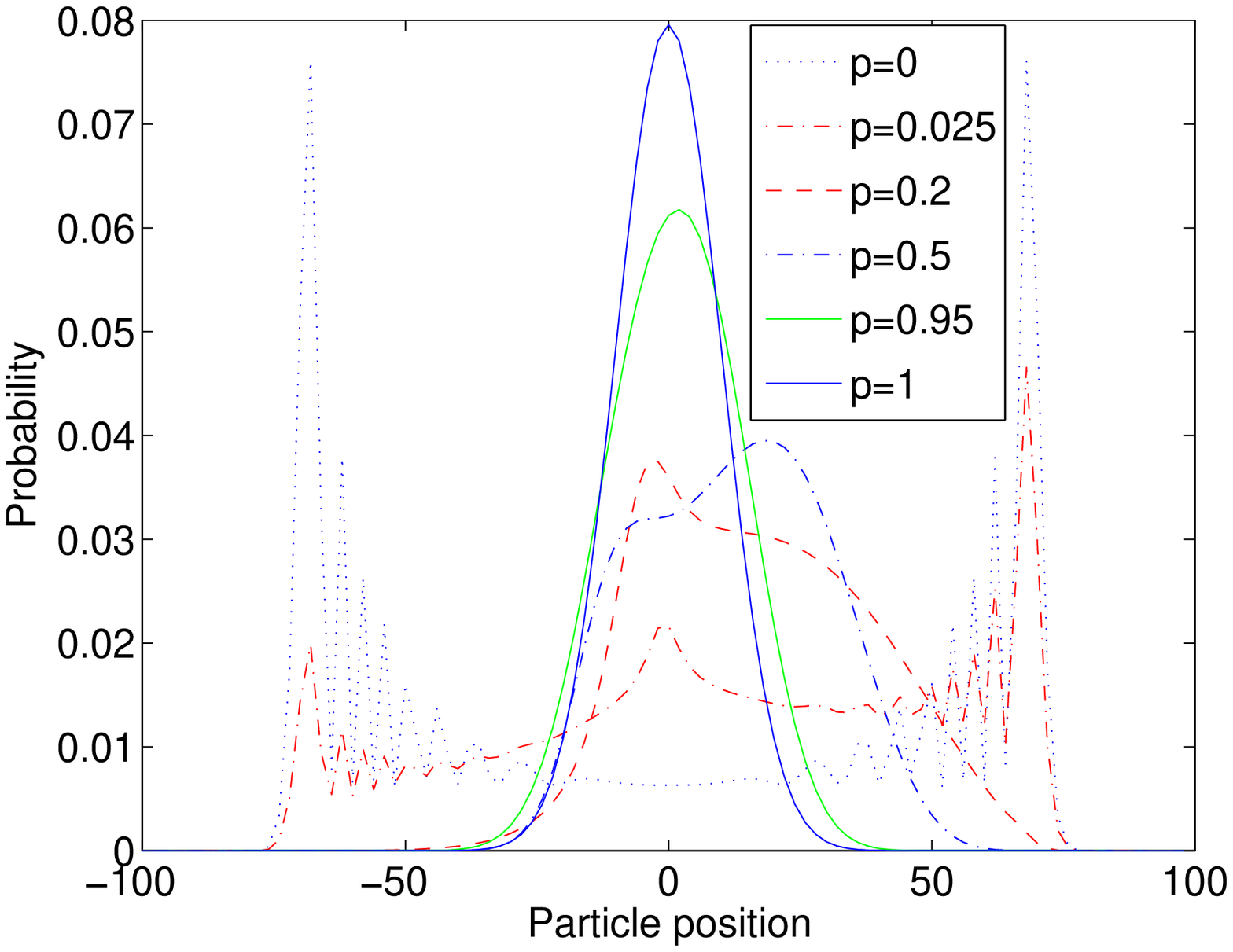}}
\subfigure[]{\includegraphics[width=8.6cm]{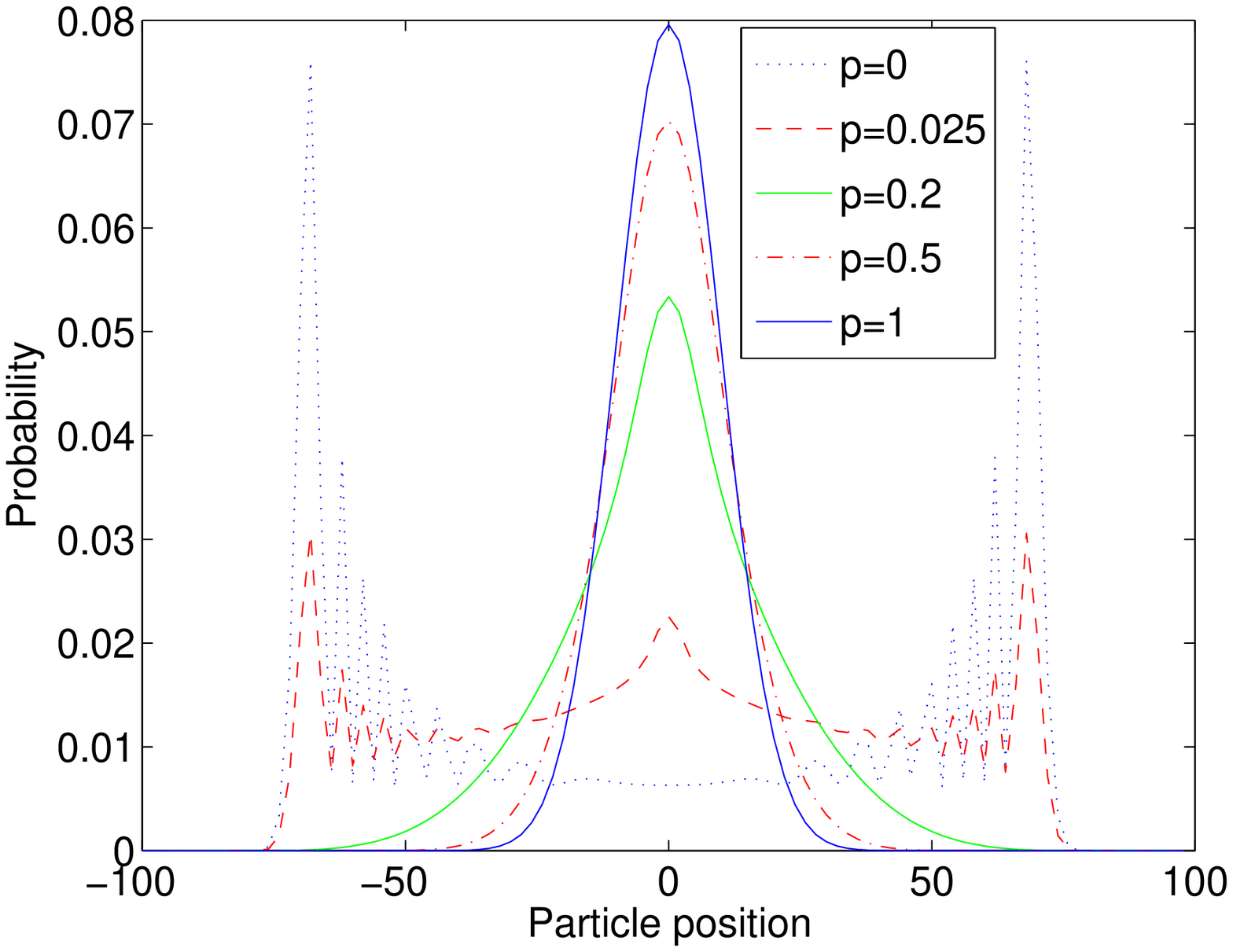}}
\caption{(color online) Onset of classicality is seen to be accentuated in
a (Hadamard) quantum walk subjected
to generalized amplitude damping with increasing temperatures.
Figure (\ref{fig:ampdamp}(a)) depicts the $T=0$ case
($\chi=1$ in Eq. (\ref{eq:bma2_lambda})). 
(a) Finite temperature corresponding to $\chi=0.75$ 
(b) $T = \infty$, corresponding to $\chi=0.5$. 
It may be noted that even at $T=\infty$, for 
sufficiently small coupling the distribution remains non-classical.}
\label{fig:ampdamp12}
\end{figure}

\begin{figure}
\includegraphics[width=8.6cm]{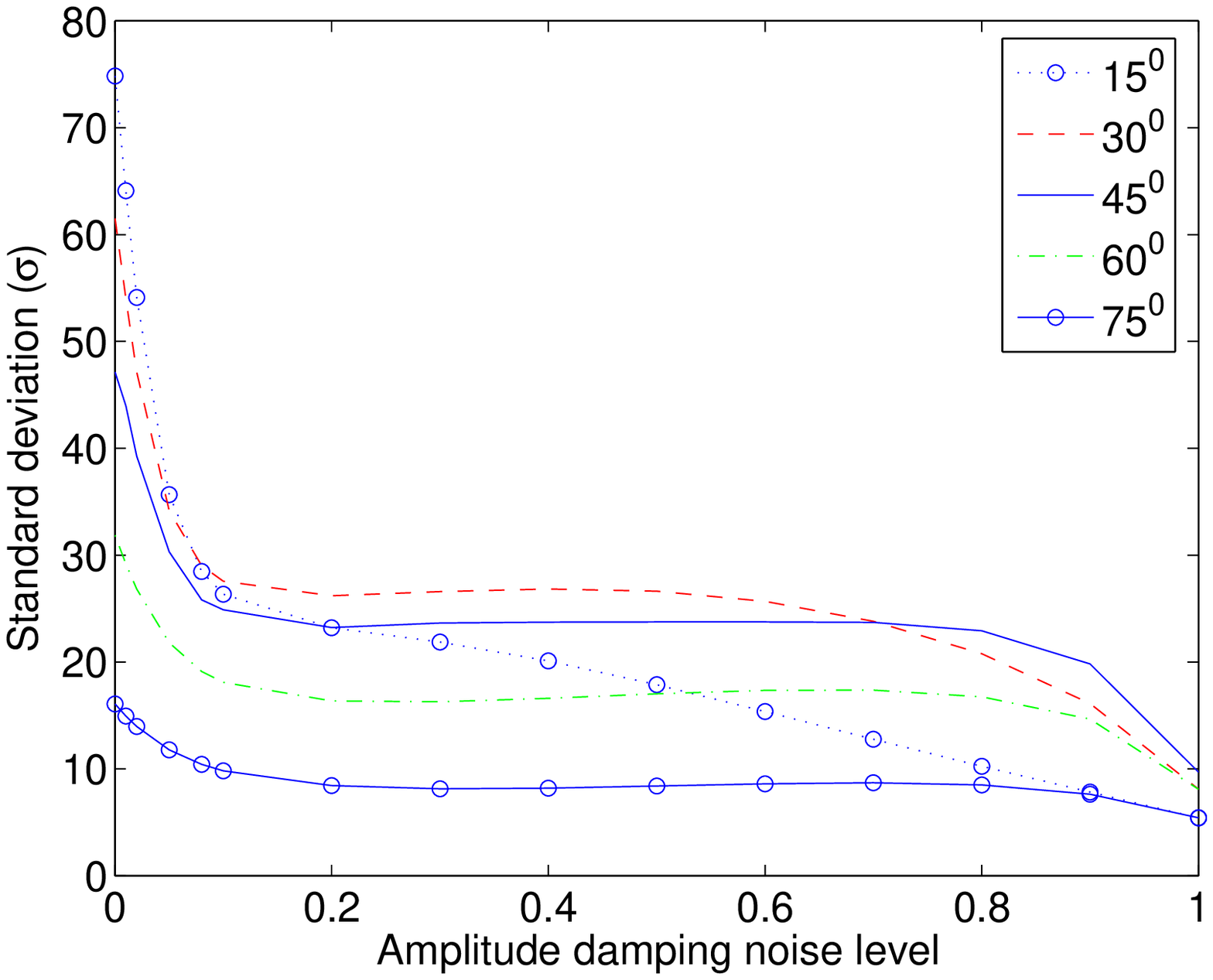}
\caption{(color online) Variation of standard  deviation with amplitude damping noise
level for various value of $\theta$, $15^{\circ}$, $30^{\circ}$, 
$45^{\circ}$, $60^{\circ}$ and $75^{\circ}$. Note that the
standard deviation for complementary
angles converge to the same value.}
\label{fig:ampdamp5}
\end{figure}

Representing  the  walker   distribution  by  its  standard  deviation
$\sigma$, we  may describe symmetry  by the ratio of  $\sigma$ without
the  symmetry  operation  to  $\sigma$ with  the  symmetry  operation.
Figure  \ref{fig:sd2} depicts  the  symmetry operation  ${\bf X}$  for
various  bit  flip  noise   levels.  The  convergence  of  the  curves
representing  various  $\theta$'s is  a  consequence  of the  complete
randomization of the measured  bit outcome in the computational basis.
This implies that  although {\bf X} is not a  symmetry of biased walk,
it does  become one in the  fully classical limit. On  the other hand,
the  symmetries ${\bf P  R X}$  remain unaffected  by noise.   We note
that, since the quantum walk  here is evolved from the symmetric state
$|0\rangle +  i|1\rangle$, and the bit  flip and phase  flip noise are
not  partial  to  the   state  $|0\rangle$  or  $|1\rangle$,  this  is
equivalent to setting  ${\bf P}$ to $1$, which  explains the fact that
the distributions  in Figures \ref{fig:env60}  and \ref{fig:env30} are
spatially symmetric.  Thus, ${\bf R  X}$ by itself becomes  a symmetry
operation,   which  is  manifested   in  the   fact  that   in  Figure
\ref{fig:sd2} the values of the curve for complementary angles are the
inverse of each other.  Figure  \ref{fig:sd3} shows that for either of
the two noises, ${\bf Z}$ is a walk symmetry.

Figure \ref{fig:sd5} depicts the symmetry of the ${\bf R X}$ operation
at all  phase flip noise  levels, as evident  from that fact  that the
values of the  curve for complementary angles are  the inverse of each
other.  From  Figures \ref{fig:sd2}, \ref{fig:sd3}  and \ref{fig:sd5},
we  note that  for the  Hadamard walk,  all three  symmetry operations
${\bf Z}, {\bf  X}$ and ${\bf R}$ are  individually preserved. This is
expected because  here ${\bf P}=1$  as stated earlier, ${\bf  R}=1$ by
definition, so that the symmetry of ${\bf PRX}$ implies ${\bf X}=1$.

With the inclusion of bias and an initial
arbitrary state, the full symmetries 
${\bf Z}$ and ${\bf P R X}$ would be required, as proved by the
following theorem.
\begin{thm}
The operations ${\bf PRX}$ and ${\bf Z}$
are symmetries for the phase-flip and bit-flip
channels.
\label{thm:phasebit}
\end{thm}
{\bf Proof.} We may look upon the phase flip channel 
(\ref{eq:phaseflip}) as a probabilistic
mixture (in the discretized walk model) of $2^n$ quantum trajectories
with $\Pi \in \{\mathbb{I}, Z\}$. 
By virtue of Theorem \ref{thm:mix}, it suffices to show that any given
unravelling is invariant under ${\bf Z}$ and ${\bf PRX}$.
Consider an unravelling $\widehat{X}^1 \equiv \cdots(IUB)(ZUB)(IUB)\cdots
= \cdots(UB)(ZUB)(UB)\cdots$.
This is the same as: $\cdots(UB)(UB^{\prime})(UB)\cdots$, where
$B^{\prime} = ZB$, noting that $Z$ commutes with $U$. Now,
${\bf Z}\widehat{X}^1 = \cdots(IZUB)(ZZUB)(IZUB)\cdots 
= \cdots(ZUB)(ZUB^{\prime})(ZUB)\cdots$, which, by Theorem \ref{thm:qutraj},
is equivalent to $\widehat{X}^1$.
Now, ${\bf PRX}\widehat{X}^1 
= \cdots(PRX~UB)(Z~PRX~UB)(PRX~UB)\cdots  
= \cdots(PRX~UB)(PR~ZX~UB)(PRX~UB)\cdots  
= \cdots(PRX~UB)(PR~ZXZ~ZUB)(PRX~UB)\cdots  
= \cdots(PRX~UB)(PR~(-X)~UB^{\prime})(PRX~UB)\cdots$, 
which, by Theorem \ref{thm:qutraj},
is equivalent to $\widehat{X}^1$, since an overall 
phase factor of $\pm 1$
is irrelevant. Thus, the phase-flip channel is symmetric
w.r.t. the operations ${\bf Z}$ and ${\bf PRX}$.

Regarding the bit-flip channel (\ref{eq:bitflip}):
as in the above case, consider an unravelling 
$\widehat{X}^2 \equiv \cdots(UB)(XUB)(UB)\cdots$.
This is the same as: $\cdots(UB)(U^{\dag}B^{\prime\prime})(UB)\cdots$, where,
as may be seen by direct calculation,
$B^{\prime\prime} = XB$. Now,
${\bf Z}\widehat{X}^2  
= \cdots(Z~UB)(X~Z~UB)(Z~UB)\cdots
= \cdots(Z~UB)(X~ZX~XUB)(Z~UB)\cdots
= \cdots(Z~UB)((-Z)~U^{\dag}B^{\prime\prime})(Z~UB)\cdots$,
which, by Theorem \ref{thm:qutraj},
is equivalent to $\widehat{X}^2$, since an overall 
phase factor of $\pm 1$
is irrelevant. 

Further,
${\bf PRX}\widehat{X}^2  
= \cdots(PRX~UB)(X~PRX~UB)(PRX~UB)\cdots
= \cdots(PRX~UB)(PRX~X~UB)(PRX~UB)\cdots
= \cdots(PRX~UB)(PRX~U^{\dag}B^{\prime\prime})(PRX~UB)\cdots$,
which, by Theorem \ref{thm:qutraj},
is equivalent to $\widehat{X}^2$.
\hfill $\blacksquare$
\bigskip

\subsection{Decoherence via generalized
amplitude damping channel \label{sec:envb}}

Here we study the behaviour of quantum walk subjected to a generalized
amplitude damping (with temperature
$T \ge 0$), which would reduce at $T=0$ to the
amplitude damping channel.
As an example of a physical process that realizes the generalized
amplitude damping channel, we consider a two-level system interacting
with a reservoir of harmonic oscillators, with the system-reservoir
interaction being dissipative
and of the weak Born-Markov type \cite{bp02,srienv06} leading to a standard
Lindblad equation, which in the interaction
picture has the following form \cite{srib06} 
\begin{equation}
\frac{d}{dt}\rho^s(t) = \sum_{j=1}^2\left(
2R_j\rho^s R^{\dag}_j - R_j^{\dag}R_j\rho^s - \rho^s R_j^{\dag}R_j\right),
\end{equation}
where $R_1 = (\gamma_0(N_{\rm th}+1)/2)^{1/2}R$,
$R_2 = (\gamma_0N_{\rm th}/2)^{1/2}R^{\dag}$ and 
$N_{\rm th} = (\exp(\hbar\omega/k_B T) - 1)^{-1}$,
is the Planck distribution giving the number of thermal
photons at the frequency $\omega$, and $\gamma_0$ is the
system-environment coupling constant. Here
$R = \sigma_-\cosh(r) + e^{i\Phi}\sigma_+\sinh(r)$, and the quantities
$r$ and $\Phi$ are the environmental squeezing parameters and
$\sigma_{\pm} = \frac{1}{2}\left(\sigma_1 \pm i\sigma_2\right)$.
For the generalized amplitude damping channel, we set $r = \Phi = 0$.
If $T=0$, so that $N_{\rm th}=0$, then $R_2$ vanishes, and a single
Lindblad operator suffices.

The generalized amplitude damping channel is characterized by
the following Kraus operators \cite{srib06}:
\begin{eqnarray}
\label{eq:gbmakraus}
\begin{array}{ll}
E_0 \equiv \sqrt{\chi}\left[\begin{array}{ll} 1 & 0 \\ 0 & 
\sqrt{1-p(t)}
\end{array}\right]; ~~~~ &
E_1 \equiv \sqrt{\chi}\left[\begin{array}{ll} 0 & \sqrt{p(t)} \\ 0 & 0
\end{array}\right], \\
E_2 \equiv \sqrt{1-\chi}\left[\begin{array}{ll} 
\sqrt{1-p(t)} & 0 \\ 0 & 1
\end{array}\right]; ~~~~ &
E_3 \equiv \sqrt{1-\chi}\left[\begin{array}{ll} 0 & 0 \\ \sqrt{p(t)} & 0
\end{array}\right],  
\end{array}
\end{eqnarray} 
where 
\begin{equation}
\label{eq:bma2_lambda}
p(t) \equiv 1 - e^{-\gamma_0(2N_{\rm th} +1) t};\hspace{1.0cm} 
\chi \equiv \frac{1}{2}\left[1 + \frac{1}{2N_{\rm th} +1}\right].
\end{equation}
When $T = 0$, $\chi=1$, and for $T \rightarrow \infty$, $\chi=1/2$.

The density operator at a future time can be obtained as
\cite{srib06}
\begin{equation}
\label{eq:bmrhos}
\rho^s (t) = \begin{pmatrix} {\frac{1}{2}} (1 + A_1) & A_2
\cr A_2^*  & 
{\frac {1}{ 2}} (1 - A_1)
\end{pmatrix}, 
\end{equation}
where
\begin{equation}
A_1 \equiv \langle\sigma_3(t)\rangle
= e^{-\gamma_0 (2N_{\rm th} + 1)t} \langle 
\sigma_3 (0) \rangle - {\frac{1}{(2N_{\rm th} + 1)}} 
\left(1 - e^{-\gamma_0 (2N_{\rm th} + 1)t} 
\right), \label{4m} 
\end{equation}
\begin{equation}
A_2 = e^{-{\frac{\gamma_0} {2}}(2N_{\rm th} + 1)t}
\langle \sigma_- (0) \rangle \rangle. \label{4n}
\end{equation}
Figures \ref{fig:ampdamp}, \ref{fig:ampdamp34} and \ref{fig:ampdamp12}
depict the onset of classicality with increasing coupling strength
(related to $p$) and temperature (coming from $\chi$).
Figure \ref{fig:ampdamp}, which shows the effect of an amplitude
damping channel on a Hadamard walker at zero temperature, illustrates
the breakdown of ${\bf R X}$ symmetry even though the initial
state is $|0\rangle + i|1\rangle$. This is because,
in contrast to the phase-flip
and bit-flip channels, the generalized amplitude damping is not
symmetric towards the states $|0\rangle$ and $|1\rangle$.
However, the extended symmetry ${\bf PRX}$ is preserved both for
Hadamard as well as biased walks, as seen
from Figures \ref{fig:ampdamp} and \ref{fig:ampdamp34}, respectively.

From  Figures (\ref{fig:ampdamp}(a)),  (\ref{fig:ampdamp12}(a,b)), the
onset  of classicality  with increasing  temperature is  clearly seen.
Figure \ref{fig:ampdamp5} presents  the standard deviation for quantum
walks on  a line with  various biases, subjected to  amplitude damping
noise.  The  standard  deviation  for  complementary  angles  ($\theta
\leftrightarrow \pi/2-\theta$)  is seen to converge to  the same value
in   the  fully   classical  limit.    This  may   be   understood  as
follows. First,  we note that since  ${\bf PRX}$ is a  symmetry of the
quantum walk,  and the  effect of ${\bf  P}$ does  not show up  in the
standard  deviation  plots,  ${\bf  RX}$  by  itself  is  an  apparent
symmetry.   Further, in  the classical  limit the  measurement outcome
being  a  unique  asymptotic  state for  the  (generalized)  amplitude
damping channel, effectively ${\bf X} \simeq 1$, which makes ${\bf R}$
a symmetry operation.

The following theorem generalizes  Theorem \ref{thm:qutraj} to an open
system subjected to a generalized amplitude damping channel.
\begin{thm}
The  operations ${\bf  Z}$  and  ${\bf PRX}$  are  symmetries for  the
generalized amplitude damping channel.
\label{thm:ampdamp}
\end{thm}
{\bf Proof.} By  virtue of Theorem \ref{thm:mix}, it  suffices to show
that  any given  unravelling is  invariant under  ${\bf Z}$  and ${\bf
PRX}$. Consider an unravelling
\begin{subequations}
\begin{eqnarray}
\label{eq:b} 
\widehat{X}^3 &\equiv& \cdots (E_0 UB)(E_1 UB)(E_2 UB)(E_3 UB)\cdots
 \\
&\equiv& \cdots (UB_{(0)})(U^{\dag}B_{(1)})(UB_{(2)})(U^{\dag}B_{(3)})
\cdots, \label{eq:b2}
\end{eqnarray}
\end{subequations}
where the non-unitary matrices are given by $B_{(j)} = E_jB$.
Now, 
\begin{eqnarray}
\label{eq:zb}
{\bf Z}\widehat{X}^3 
&=&  \cdots(E_0 Z UB)(E_1 ZUB)(E_2 ZUB)(E_3 ZUB)\cdots \nonumber \\
&=& \cdots (Z E_0 UB)((-Z) E_1 UB)(Z E_2 UB)((-Z)E_3 UB)\cdots \nonumber \\
&=& \cdots(Z UB_{(0)})((-Z) U^{\dag}B_{(1)})
(ZUB_{(2)})((-Z) U^{\dag}B_{(3)})
\cdots \nonumber \\
&=& \cdots(UB_{(0)}^{(1)})(-U^{\dag}B_{(1)}^{(1)})
(UB_{(2)}^{(1)})(-U^{\dag}B_{(3)}^{(1)}) \cdots 
\end{eqnarray}
Ignoring  the  overall  $\pm  1$  factor  in  Eq.  (\ref{eq:zb}),  and
comparing  it  with  Eq.   (\ref{eq:b}),  and  noting  that  that  the
derivation of  the proof of  Theorem \ref{thm:qutraj} did  not require
the matrices  $B_j$ to  be unitary, we  find along similar  lines that
${\bf Z}\widehat{X}^3$ is equivalent to $\widehat{X}^3$.

The following may be directly verified
\begin{subequations}
\label{eq:bprx}
\begin{eqnarray}
{\bf PRX}\widehat{X}^3 
&=& \cdots(E_0 PRX UB)(E_1 PRX UB)(E_2 PRX UB)(E_3 PRX UB) \cdots \\
&=& \cdots(E_0 UB^{(2)})(E_1 UB^{(2)})(E_2 UB^{(2)})(E_3 UB^{(2)}) 
\cdots \label{eq:bprx2} \\
&=& \cdots(UB_{(0)}^{(2)})(U^{\dag}B_{(1)}^{(2)})(UB_{(2)}^{(2)})
 (U^{\dag}B_{(3)}^{(2)}) \cdots, \label{eq:bprx3}
\end{eqnarray}
\end{subequations}
which,    by    Theorem    \ref{thm:qutraj},    is    equivalent    to
$\widehat{X}^3$. For  proof of Eq. (\ref{eq:bprx2}), see  the proof of
Theorem \ref{thm:bias}.  Eq. (\ref{eq:bprx3}) is  obtained analogously
to Eq. (\ref{eq:b2}), except that the matrix $B^{(2)}$ is used instead
of $B$.  \hfill $\blacksquare$
\bigskip

This may be expressed by the statement 
\begin{subequations}
\begin{eqnarray}
\label{eq:symn}
{\cal N}\widehat{W} &\simeq& {\cal N}{\bf Z}\widehat{W}, \\
{\cal N}\widehat{W} &\simeq& {\cal N}{\bf PRX}\widehat{W}, 
\end{eqnarray}
\end{subequations}
which generalizes Eq. (\ref{eq:symm}).
These results show that the symmetries persist
for dephasing (phase flip), bit flip and (generalized)
amplitude damping channels.

\section{Quantum walk on a cycle \label{sec:cycle}}

In this work, though we  are primarily concerned with symmetries for a
quantum walk on a line, and the influence of noise on them, we briefly
consider in  this Section an extension  of the above  ideas to quantum
walks on a cycle. Further extensions  would be quantum walks on a more
general  graph  \cite{vk06,om06}  or  in  higher dimensions  $d  >  2$
\cite{mac02}. In the former, the  1D walk is generalized to $N$-cycles
and to  hypercubes, including  the effect of  phase noise in  the coin
space, and decoherence in position space.  In the latter, the Hadamard
transformation is  generalized to  a non-entangling tensor  product of
Hadamards,  or to  an  entangling discrete  Fourier  transform or  the
Grover  operator.  They  bring in  many novel  features absent  in the
quantum walk  on a line. Here  we will restrict  ourselves to pointing
out that quantum  walk on a cycle differs considerably  from walk on a
line, both with respect to symmetry operations as well as noise.

In contrast to the case of quantum walk on a line,
none of the four discrete symmetries of Theorem \ref{thm:bias}
hold in general for unitary quantum walk on a cycle or closed path. Thus, 
if $B$ in Eq. (\ref{eq:bias1}) is replaced by any of $B^{(1)}$, 
$B^{(2)}$, $B^{(3)}$, or $B^{(4)}$, given by Eq. (\ref{eq:bias20}), 
the spatial probability distribution is not guaranteed to be the same. 

\begin{thm} The operation $G: B \rightarrow B^{\star}$ 
is in general not a symmetry of the quantum walk on a cycle.
\end{thm}
\noindent {\bf Proof.} 
For the cyclic case, in place of Eq. (\ref{eq:bitobit}), we now have
\begin{subequations}
\begin{eqnarray}
\label{eq:cybitobit}
|\Psi_1\rangle &=& 
(UB)^n|\alpha,\beta\rangle = \sum_{j_1,j_2,\cdots,j_n}
b_{j_n,j_{n-1}}\cdots b_{j_2,j_1}b_{j_1,\alpha}
|j_n,\beta+2J - n~({\rm mod}~ R)\rangle, \\
|\Psi_2\rangle &=& 
(UB^{(1)})^n|\alpha,\beta\rangle = \sum_{j_1,j_2,\cdots,j_n}
b_{j_n,j_{n-1}}\cdots b_{j_2,j_1}b_{j_1,\alpha}
(e^{i\phi})^{j_{n-1}+ \cdots + j_1 + \alpha}
|j_n,\beta+2J - n ~({\rm mod}~ R)\rangle, 
\end{eqnarray}
\end{subequations}
where $R$ is the number of sites in the cycle. 
For  an    arbitrary     state     $|a,b\rangle$    in     the
computational-and-position  basis, 
we have
\begin{subequations}
\begin{eqnarray}
\langle  a,b|\Psi_1\rangle &=& \sum_{j_1,j_2,\cdots,j_{n-1} \in {\cal J}}
b_{a,j_{n-1}}\cdots b_{j_2,j_1}b_{j_1,\alpha} \label{eq:ny1} \\
\langle  a,b|\Psi_2\rangle &=& \sum_{j_1,j_2,\cdots,j_{n-1} \in {\cal J}}
b_{a,j_{n-1}}\cdots b_{j_2,j_1}b_{j_1,\alpha}
(e^{i\phi})^{j_{n-1}+ \cdots + j_1 + \alpha} \nonumber \\
&\equiv& \sum_{j_1,j_2,\cdots,j_{n-1} \in {\cal J}} b_{a,j_{n-1}}\cdots 
b_{j_2,j_1}b_{j_1,\alpha}(e^{i\phi\epsilon}), \label{eq:ny2}
\end{eqnarray}
\end{subequations}
where   ${\cal    J}$   is   the   set    of   binary   $(n-1)$-tuples
$j_1,j_2,\cdots,j_{n-1}$ such that $J = j_1 + j_2 + \cdots + j_{n-1} +
a$  satisfies $b  =  \beta +  2J-n  ~({\rm mod}~  R)$.   We find  that
$\epsilon = \alpha - a + J$ and $J = (b + n - \beta)/2 + mR$, where $m
=  0,1,2,\cdots,  \lfloor  n/R   \rfloor$.  Thus,  the  terms  in  the
superposition (\ref{eq:ny1})  are not in general  identical with those
in (\ref{eq:ny2}),  apart from a common factor,  unless $\phi=0, 2\pi,
4\pi,\cdots$.  A  similar argument  can be used  to show that the terms in 
$\langle  \overline{a},b|\Psi_1\rangle$ are not in general
the  same as those  in $\langle  \overline{a},b|\Psi_2\rangle$.  Given
the     independence    of     $\phi$     from    the     coefficients
$b_{j_2,j_1}b_{j_1,\alpha}$,  it  is   not  necessary  that  $|\langle
a,b|\Psi_1\rangle|^2  +   |\langle  \overline{a},b|\Psi_1\rangle|^2  =
|\langle           a,b|\Psi_2\rangle|^2           +           |\langle
\overline{a},b|\Psi_2\rangle|^2$.  The equality  holds in general (for
arbitrary unitary  matrix $B$ and time  parameter $n$) if  and only if
$\phi=0, 2\pi,  4\pi, \cdots$.  Repeating the  argument for $B^{(2)}$,
$B^{(3)}$ and $B^{(4)}$, we find that all the four discrete symmetries
of   Theorem   \ref{thm:bias}   break   down   in   general.    \hfill
$\blacksquare$
\bigskip

We  note  that for  phase  flip  symmetry,  where $e^{i\phi}=-1$,  the
superposition   terms   in  Eq.  (\ref{eq:ny2}) may differ  from   the
corresponding  terms  in  Eq.   (\ref{eq:ny1}) only  with  respect  to
sign. Given  that all the terms  like $b_{j_2,j_1}$, $b_{j_1,\alpha}$,
etc.  are built from  a small  set of  trignometric functions of the
three parameters  $\theta$, $\zeta$ and  $\xi$, certain values  of $n$
may  render   the  right  hand  sides  of   Eqs.   (\ref{eq:ny1})  and
(\ref{eq:ny2}) equal. However in  general, this equality will not hold
for arbitrary $n$.
\begin{figure}
\begin{center}
\subfigure[]{\includegraphics[width=8.6cm]{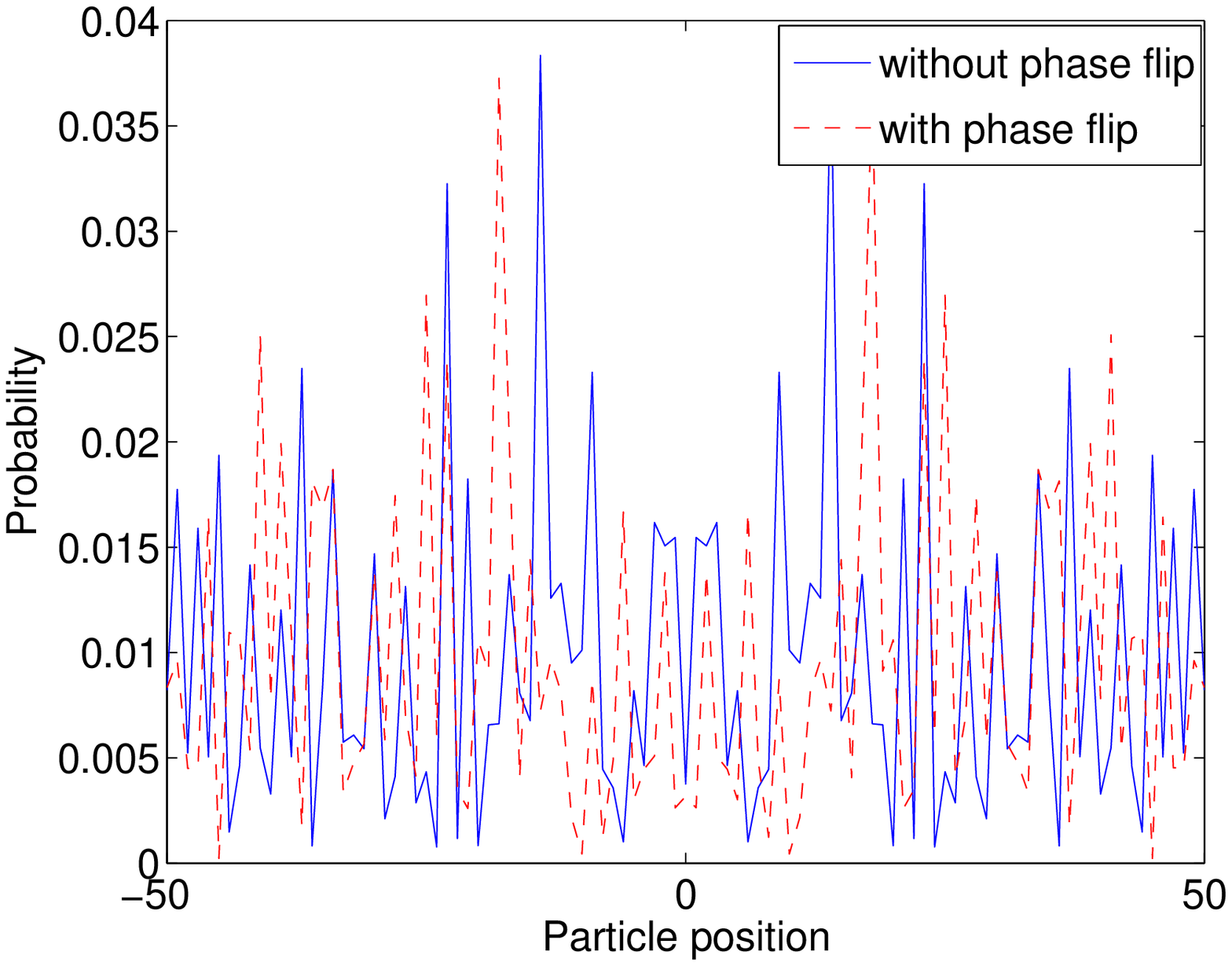}}
\hfill
\subfigure[]{\includegraphics[width=8.6cm]{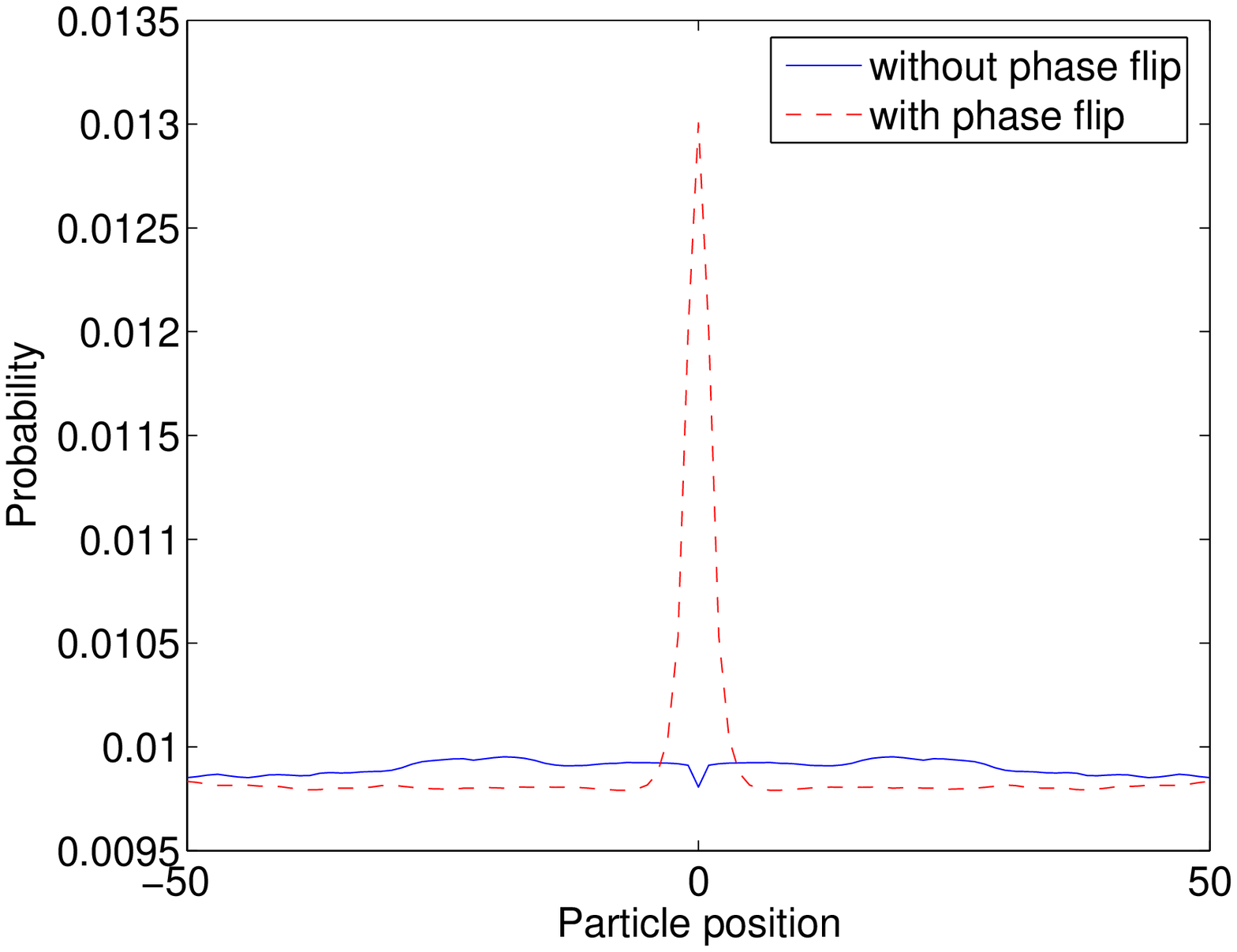}}
\caption{(color online) An instance of breakdown of phase flip symmetry in a 
unitary quantum walk on a cycle, where the two extreme points 
(located at positions $\pm 50$) on
the plot are spatially adjacent. The number of sites is 101 and 
$t=5000$ (in units of discrete time-steps), 
with bias angle $\theta=30^{\circ}$. 
(a) The solid curve represents the positional probability distribution
without any symmetry operation applied, while the dashed curve
represents that with a phase flip operation applied at each 
walk step. (b) The same as the above, but with time-averaging applied over
every 50 steps, in order to more clearly bring out the breakdown in
symmetry.}
\label{fig:withoutnoise} 
\end{center}
\end{figure}

\begin{figure}
\begin{center}
\includegraphics[width=8.6cm]{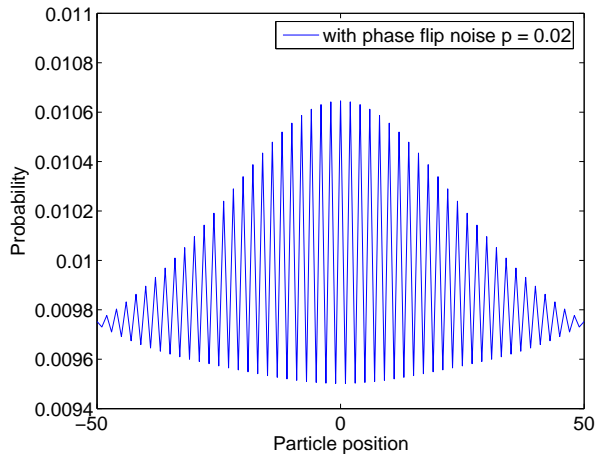}
\caption{(color online) Restoration of phase flip symmetry in a noisy quantum walk on
a  cycle, where  the  two extreme  points  on the  plot are  spatially
adjacent. The number of sites is 101 and $t=5000$ (in units of discrete
time-steps),
with bias angle $\theta=30^{\circ}$.   The figure depicts the position
probability distribution with or without phase flip symmetry operation
applied  at  each  step,  with  phase  damping  noise  level  $p=0.02$
[Eq. (\ref{eq:phaseflip})].  After sufficiently long time, the quantum
walk  reaches the  uniform distribution,  typical of  classical random
walk.}
\label{fig:withnoise} 
\end{center}
\end{figure}

An instance of breakdown of phase flip 
symmetry in the unitary quantum  walk on  a  cycle is
demonstrated  in the example  of Figure  \ref{fig:withoutnoise}.
The  profile of  the
position probability  distribution varies  depending on the  number of
sites and  the evolution time.  Remarkably, this  symmetry is restored
above   a  threshold   value  of   noise.   The   pattern   in  Figure
\ref{fig:withnoise} corresponds  to phase noise  with $p=0.02$ applied
to  a quantum  walk,  either with  or  without a  phase flip  symmetry
operation.   We   note  that  the  introduction  of   noise  tends  to
classicalize the random walk, hence causing it to asymptotically reach
a  uniform  distribution  \cite{vk06,om06}.  The  above-mentioned  symmetry
restoration happens  well before the uniformity sets  in.  The initial
lack of symmetry  gradually transitions to full symmetry  as the noise
level is increased.   Thus, the role of symmetry  operations and noise
is quite different  in the case of quantum walk  on cycles as compared
with that on a line.

A more detailed treatment of symmetries and noise in a quantum walk on 
a cycle and general graphs of other topologies will be presented 
elsewhere \cite{new}.

\section{Experimental realization in physical systems}
\label{sec:qwbec}

Experimental  realization of quantum  walk using  any of  the proposed
schemes is  not free  from noise due  to environmental  conditions and
instrumental interference.  In particular,  noise can be a major issue
in the scaling  up of the number of steps  in already realized quantum
walk systems.  Understanding the symmetries of the noisy and noiseless
quantum walk could  greatly help in the improvement  of known
techniques  and in further  exploration of other possible  systems
where quantum walk can be realized  on a large scale.  In this section
we  discuss the  realization  in a  nuclear  magnetic resonance  (NMR)
quantum-information processor and ultra-cold atomic systems.

\subsection{NMR quantum-information processor}
\label{subsec:nmr}

Continuous time  \cite{du} and discrete time  \cite{ryan} quantum walk
have  been successfully  implemented in  a nuclear  magnetic resonance
(NMR) quantum-information  processor. Considering the  benefits of the
effect  of   decoherence  on   the  quantum  walk   \cite{viv,  brun},
Ref.  \cite{ryan} has  also  experimentally added  decoherence on  the
discrete time quantum walk by implementing dephasing in NMR. By adding
decoherence after each step they  have shown the transition of quantum
walk to the classical random walk.

In NMR spectroscopy  of the given system (molecule)  the extent of its
isolation from  the environment  is determined in  terms of  its phase
coherence time $T_{2}$ and its  energy relaxation time $T_{1}$. If the
pulse  sequence is  applied to  the NMR  quantum-information processor
within the  time $T  < {T_{2},  T_{1}}$, the system  is free  from the
environmental effects.  The pulse  sequence exceeding the time $T_{2}$
can be considered  to affected by the dephasing  channel and the pulse
sequence exceeding the  time $T_{1}$ can be considered  to be affected
by  the  amplitude damping  channel.   In  experiments  of time  scale
greater than the time $T_{2}$  or $T_{1}$, a refocusing pulse sequence
is  applied to  compensate  for the  environmental  effects.  In  Ref.
\cite{ryan} the  pulse sequence for  the quantum walk  was implemented
within the time $T_{1}$ and $T_{2}$.

The environmental effect (noise)  on quantum walk symmetries presented
in this  article can be verified in  the NMR system by  scaling up the
number of  steps of  quantum walk realized.  By applying  a controlled
amount  of the  refocusing  pulse sequence,  the  effect of  different
levels of noise can be experimentally verified.

\subsection{Ultra-cold atoms}
\label{subsec:atoms}

There have  been various schemes suggested to  implement quantum walks
using neutral  atoms in an  optical lattice \cite{rauss,  eckert}.  In
Ref. \cite  {mandel}, the controlled coherent  transport and splitting
of  atomic wave  packets in  spin dependent  optical lattice  has been
experimentally  demonstrated  using  rubidium atoms.  A  Bose-Einstein
condensate of up to $3 \times  10^{5}$ atoms is initially created in a
harmonic  magnetic  trap.  A  three  dimensional  optical  lattice  is
superimposed  on the  Bose-Einstein  condensate and  the intensity  is
raised  in order  to drive  the system  into a  Mott  insulating phase
\cite{greiner}.

Two of the three orthogonal  standing wave light fields is operated at
one  wavelenght, $\lambda_{y,z}  = 840$  nm and  the third  along the
horizontal direction is tuned to the wavelength $\lambda _{x} =785$ nm
between the  fine structure  splitting of the  rubidium $D1$  and $D2$
transitions.    Along  this  axis   a  quarter   wave  plate   and  an
electro-optical  modulator  (EOM)  is  placed  to  allow  the  dynamic
rotation of  the polarization  vector of the  retro-reflected laser
beam through an  angle $\theta$ by applying an  appropriate voltage to
the  EOM.  After  reaching  the Mott  insulating  phase, the  harmonic
magnetic  field is completely  turned off  but a  homogeneous magnetic
field  along the  $x$ direction  is  maintained to  preserve the  spin
polarization  of  the atoms.   The  light field  in  the  $y$ and  $z$
direction  is  adiabatically  turned  off to  reduce  the  interaction
energy, which  strongly depends on the  confinement of the  atoms at a
single lattice site.

A  standing  wave  configuration  in  the $x$  direction  is  used  to
transport  the  atoms.  By  changing  the  linear polarization  vector
enclosing angle $\theta$, the separation between the two potentials is
controlled. By rotating the polarization angle $\theta$ by $\pi$, with
the  atom in  a superposition  of  internal states,  the spatial  wave
packets of the  atom in the $|0\rangle$ and  the $|1\rangle$ state are
transported  in opposite  directions.  The  final state  after  such a
movement   is   then    given   by   $1/\sqrt{2}(|0,x-1\rangle   +   i
\exp(i\beta_p)|1, x+1\rangle)$.  The phase $\beta_p$ between the separated
wave-packets depends  on the accumulated kinetic  and potential energy
phases in  the transport process and  in general will  be nonzero. The
coherence  between  the two  wave-packets  is  revealed by  absorption
imaging  of the momentum  distribution. A  $\pi/2$ microwave  pulse is
applied before  absorption imaging to erase  the which-way information
encoded in the hyperfine states.

However,  to  increase the  separation  between  the two  wave-packets
further, one could increase the polarization angle $\theta$ to integer
multiples of $180^{\circ}$. To  overcome the limitation of the maximum
voltage  that can  be applied  to  the EOM,  a $\pi$  pulse after  the
polarization  is  applied,  thereby  swapping  the roles  of  the  two
hyperfine  states.   The   single  particle  phase  $\beta_p$  remains
constant throughout  the atomic cloud and is  reproducible.  After the
absorption imaging a Gaussian  envelope of the interference pattern is
obtained.

One can build up on the  above technique to implement a quantum walk
which introduces  phase along with  each splitting (step).   The above
setup can be modified by dividing the separations into small steps and
introducing a $\pi/2$ pulse after each separation without intermediate
imaging. A phase $\beta_p$ is introduced in each step.  The absorption
imaging of the distribution of  the atomic cloud after $n$ steps would
give the interference pattern of the quantum walk.

This effect of  the addition of phase during  the quantum walk process
can be easily understood from  the phase damping channel and arbitrary
phase rotation presented in this  article. The addition of $\pi$ pulse
to overcome the  limit of EOM is a bit flip  operation in the quantum
walk.

There have  been other proposals  for physical realization  of quantum
walk using  Bose-Einstein condensate (BEC)  \cite{chandra06} where the
unitary shift  operator induces a  bit flip.  A {\em  stimulated Raman
kick}  is  used   as  a  unitary  shift  operator   to  translate  the
Bose-Einstein  condensate   in  the  Schr\"odinger  cat   state  to  a
superposition in position  space. Two selected levels of  the atoms in
the  Bose-Einstein  condensate  are   coupled  to  the  two  modes  of
counter-propagating  laser  beams.   The  stimulated  Raman  kick,  in
imparting  a translation in  position space,  also flips  the internal
state of the  Bose-Einstein condensate.  An rf pulse  ($\pi$ pulse) is
suggested as  a compensatory mechanism  to flip the internal  state of
the condensate  back to  its initial value  after every  unitary shift
operator. From  the ${\bf PRX}$  symmetry pointed in this  article, it
follows that  there is no need  for the compensatory  operation for an
unbiased quantum  walk started in the state  $|0\rangle + i|1\rangle$.
The availability of walk symmetries could also be useful for exploring
other  possible  physical  implementations  which  induce such
symmetry operations along with the translation.

In the  most widely  studied version of  quantum walk, a  quantum coin
toss (Hadamard operation) is  used after every displacement operation.
Continuous  external  operations on  a  particle  confined  in a  trap
reduces the confinement  time of the particle. Reducing  the number of
external operations will benefit the scaling up of the number of steps
of the quantum  walk. One can workout a system  where the quantum coin
toss operation  is eliminated by  transferring the burden  of evolving
the  particle   in  superposition  of   the  internal  state   to  the
displacement operator itself  \cite{chandra062, css}.  Such a transfer
may  introduce  additional  operations  that correspond  to  the  walk
symmetry, and may thus be  ignored.  Further, systems of this kind are
expected  to be affected  by amplitude  damping as  one of  the states
might be more  stable than the other one in  the trap.  The present
work could help to optimize such a noisy quantum walk.

In  a  scheme  suggested  using quantum  accelerator  mode  \cite{ma},
different  internal  states  of  an atom  receive  different  momentum
transfer with  each alternative kick, giving  different walking speeds
in the two directions and this can be seen as a biased walk. Our study
of symmetry  in a  noisy, biased quantum  walk could help  improve the
above technique and make it easier for its experimental realization.

\section{Conclusions} \label{conclusion}

Our  work considers  variants of  quantum walks  on a  line  which are
equivalent  in  the  sense   that  the  final  positional  probability
distribution  remains the  same in  each variant.   In  particular, we
consider variants  obtained by the  experimentally relevant operations
of $Z$  or $X$ applied  at each quantum  walk step, with  the symmetry
operations  given  by  ${\bf  Z}$  and ${\bf  PRX}$.   This  could  be
experimentally advantageous since  practical constraints may mean that
one of the variants is preferred over the rest.  A specific example is
the  simplification of  the  implementation of  quantum  walk using  a
Bose-Einstein condensate  with a  stimulated Raman kick  providing the
conditional translation operation.   What is especially interesting is
that these symmetries are preserved  even in the presence of noise, in
particular,  those  characterized by  the  phase  flip,  bit flip  and
generalized amplitude  damping channels. This is  important because it
means that  the equivalence of these  variants is not  affected by the
presence of noise, which would be inevitable in actual experiments.
The symmetry of the phase operation under phase noise is intuitive,
considering that this noise has a Kraus representation consisting of
operations that are symmetries of the noiseless quantum walk. However, for
the PRX symmetry under phase noise, and for any symmetry under other noisy
channels (especially in the case of generalized amplitude damping
channel), the connection was not obvious before the analysis was
completed.

Our results  are supported by  several numerical examples  obtained by
evolving  the density  operator in  the Kraus  representation. However,
analytical proofs  of the effect  of noise on symmetries  are obtained
using the quantum trajectories  approach, which we find convenient for
this situation. We  also present the quantum walk  on the cycle, which
can be  generalized to  any closed graph.   An interesting  fact that
comes out is that the symmetry  breaks down in general but is restored
above a  certain noise level. Some  representative plots demonstrating
the effect  of the  phase damping channel  on phase flip  symmetry are
presented.

Finally we have discussed the experimental realization of quantum walk
in a  physical system, such as  NMR and ultra cold  atoms, as examples
where these studies can be  beneficial for improving the efficiency of
the implementation on large scales.



\begin{thebibliography}{99}

\bibitem{barber-ninham} M. N. Barber and B. W. Ninham, 
{\it Random and Restricted walks: Theory and Applications} (Gordon and 
Breach, New York, 1970).

\bibitem{chandra} S. Chandrasekhar, Rev. Mod. Phys. {\bf 15}, 1 (1943).

\bibitem{mark}  M. Jerrum  and  A. Sinclair,  {\it The  Markov
chain  Monte Carlo  method: An  approach to  approximate  counting and
integration}, Edited by Dorit S. Hochbaum, in (Approximation Algorithm
for   NP-hard   Problems,  PWS   Publishing,   Boston),  chapter   12,
p.482-520. (1996).

\bibitem  {aharonov}  Y.  Aharonov,   L.  Davidovich  and  N.  Zagury,
Phys. Rev. A {\bf 48}, 1687, (1993).

\bibitem{kempe} J. Kempe, Contemp. Phys. {\bf 44}, 307 (2003).

\bibitem{childs}  A. M.   Childs,  R.  Cleve,  E.   Deotto,  E.  Farhi
{\em et al.}, in  {\it Proceedings  of the  35th ACM
Symposium on Theory of Computing} (ACM Press, New York, 2003), p.59.

\bibitem{shenvi}  N.   Shenvi,  J.  Kempe  and   K.  Birgitta  Whaley,
Phys. Rev. A {\bf 67}, 052307, (2003).

\bibitem{childs1} A. M.  Childs, J. Goldstone,  Phys. Rev. A  {\bf 70},
022314, (2004).

\bibitem{ambainis}  A.  Ambainis,  J.  Kempe and  A.  Rivosh,  e-print
quant-phy/0402107.

\bibitem{ryan} C. A. Ryan, M.  Laforest, J. C. Boileau, and R. Laflamme,
Phys. Rev. A {\bf 72}, 062317, (2005).

\bibitem{du} J. Du, H. Li, X. Xu, M. Shi {\em et al.}, 
Phys. Rev. A {\bf 67}, 042316, (2003).

\bibitem{travaglione} B. C. Travaglione and  G. J. Milburn, Phys. Rev. A
{\bf 65}, 032310, (2002).

\bibitem{rauss} W. Dur, R. Raussendorf, V. M. Kendon and H.J. Briegel,
Phys. Rev. A {\bf 66}, 052319, (2002).

\bibitem{eckert} K.  Eckert, J. Mompart,  G. Birkl and  M. Lewenstein,
Phys. Rev. A {\bf 72}, 012327, (2005).

\bibitem{chandra06} C. M. Chandrashekar,  Phys. Rev. A {\bf 74}, 032307 (2006).

\bibitem{ma} Z.-Y. Ma, K. Burnett, M. B. d'Arcy, and S. A. 
Gardiner, Phys. Rev. A {\bf 73}, 013401, (2006).

\bibitem{ken05} V. M. Kendon and B. C. Sanders, Phys. Rev. A {\bf 71}, 
022307 (2005).

\bibitem{bru02} T. A. Brun, {\em Am. J. of Phys.} {\bf 70}, 719 (2002).

\bibitem{nc00} M. Nielsen and I. Chuang, {\em Quantum Information and
Quantum Computation} (Cambridge University Press, 2000).

\bibitem{srib06} S. Banerjee and R. Srikanth, eprint quant-ph/0611161.

\bibitem{bp02} H.-P. Breuer and F. Petruccione,
{\it The Theory of Open Quantum Systems} (Oxford University Press, 2002).

\bibitem{srienv06} R. Srikanth and S. Banerjee, to appear in Phys. Lett. A;
eprint quant-ph/0611263. 

\bibitem{vk06} V. Kendon, eprint quant-ph/0606016; 

\bibitem{om06} O. Maloyer and V. Kendon, eprint quant-ph/0612229.

\bibitem{mac02} T. D. Mackay, S. D. Bartlett, L. T. Stephenson
and B. C. Sanders, J. Phys. A: Math. Gen {\bf 35}, 2745 (2002).

\bibitem{new} S. Banerjee, R. Srikanth and C. M. Chandrashekhar, 
under preparation.

\bibitem{viv} V. Kendon and B. Tregenna, Phys. Rev. A {\bf 67}, 042315 (2003).

\bibitem{brun}   T.A.   Brun,   H.A.   Carteret,  and   A.   Ambainis,
Phys. Rev. Lett. {\bf 91}, 130602 (2003).

\bibitem{mandel} O.  Mandel, M. Greiner, A.  Widera, T. Rom {\em et al.},
Phys. Rev. Lett.  {\bf 91}, 010407 (2003).

\bibitem{greiner} M. Greiner, O. Mandel, T. Esslinger, T.W. H\"ansch,
{\em  et al.}, Nature (London) {\bf 415}, 39 (2002).

\bibitem{chandra062} C. M. Chandrashekar, eprint quant-ph/0609113.

\bibitem{css} C. M. Chandrashekar, S. Banerjee and R. Srikanth, 
under preparation.
\end{thebibliography}
\end{document}